\documentclass[10pt, a4paper, copyright, goog]{google}

\usepackage[numbers, compress]{natbib}
\usepackage{xurl}
\usepackage{hyperref}
\usepackage{cleveref}
\usepackage{multirow}
\usepackage{bm}
\usepackage{quoting}
\usepackage{ragged2e}
\usepackage{array}
\usepackage{enumitem}

\crefformat{footnote}{#2\footnotemark[#1]#3}

\newcolumntype{C}[1]{>{\centering\let\newline\\\arraybackslash\hspace{0pt}}m{#1}}

\usepackage{booktabs}      %
\usepackage{array}         %
\usepackage{caption}       %
\usepackage{graphicx}
\usepackage{subcaption}
\usepackage{chngcntr}
\usepackage{longtable}
\usepackage{listings}
\usepackage{xcolor}
\usepackage{twemojis}
\usepackage{CJKutf8}
\usepackage[normalem]{ulem}
\usepackage{float}

\definecolor{backcolour}{rgb}{0.95,0.95,0.92}

\lstdefinestyle{monospacedgrey}{
    backgroundcolor=\color{backcolour},
    basicstyle=\ttfamily\scriptsize,
    numbers=none,
    breaklines=true,
    breakatwhitespace=true,
    postbreak=\mbox{\textcolor{red}{$\hookrightarrow$}\space}
}

\newcommand{\fixedemoji}[1]{\resizebox{!}{1.65ex}{\twemoji{#1}}}
\newcommand{\messageedit}[1]{\textcolor{blue}{\uline{#1}}}

\paperurl{goo.gle/LearnLM-Nov25}

\uselogo{} 

\title{AI tutoring can safely and effectively support students: An exploratory RCT in UK classrooms}

\correspondingauthor{learnlm-tech-report@google.com}

\reportnumber{} %

\renewcommand{\today}{2025-11-11}

\author[]{LearnLM Team, Google \& Eedi}

\begin{abstract}
One-to-one tutoring is widely considered the gold standard for personalized education, yet it remains prohibitively expensive to scale. To evaluate whether generative AI might help expand access to this resource, we conducted an exploratory randomized controlled trial (RCT) with $\bm{N = 165}$ students across five UK secondary schools. We integrated LearnLM—a generative AI model fine-tuned for pedagogy—into chat-based tutoring sessions on the Eedi mathematics platform.
In the RCT, expert tutors directly supervised LearnLM, with the remit to revise each message it drafted until they would be satisfied sending it themselves.
LearnLM proved to be a reliable source of pedagogical instruction, with supervising tutors approving 76.4\% of its drafted messages making zero or minimal edits (i.e., changing only one or two characters).
This translated into effective tutoring support: students guided by LearnLM performed at least as well as students chatting with human tutors on each learning outcome we measured. In fact, students who received support from LearnLM were 5.5 percentage points more likely to solve novel problems on subsequent topics (with a success rate of 66.2\%) than those who received tutoring from human tutors alone (rate of 60.7\%). In interviews, tutors highlighted LearnLM's strength at drafting Socratic questions that encouraged deeper reflection from students, with multiple tutors even reporting that they learned new pedagogical practices from the model.
Overall, our results suggest that pedagogically fine-tuned AI tutoring systems may play a promising role in
delivering effective, individualized learning support at scale.
\end{abstract}

\begin{document}

\maketitle

\section{Introduction}

One-to-one tutoring is the gold standard for supporting students' learning and education. Decades of research demonstrate that individualized tutoring results in substantial gains in learning~\citep{bloom19842, kraft2024impacts, nickow2020impressive}. Unfortunately, the high cost of one-to-one tutoring and relative scarcity of educators
makes this support inaccessible for most students and classrooms. The tension between tutoring's effectiveness and inaccessibility presents an enduring challenge for education systems: can educators deliver individualized support in a way that is both highly effective and broadly scalable?

A growing number of researchers and practitioners now look to generative AI (``genAI'') as a potential solution to this challenge~\citep{kasneci2023chatgpt, khan2024brave, mollick2024co, mote2025artificial}. Indeed, a wave of new tutoring systems incorporate genAI for direct interactivity with students~\citep{zerkouk2025comprehensive}. Yet rigorous, in-classroom research on the learning efficacy of genAI remains scarce~\citep{weidlich2025chatgpt}. The evidence that does exist is mixed: while some studies suggest genAI can offer effective instruction~\citep{de2025chalkboards, kestin2025ai, pardos2024chatgpt, wang2024tutor}, others find that deploying genAI tutoring systems without appropriate pedagogical safeguards can actively harm learning~\citep{bastani2025generative, kosmyna2025your}.

Here we report the results of an exploratory randomized controlled trial (RCT) with $N = 165$ students, designed specifically to evaluate if an AI tutor can safely and effectively support students in UK secondary school classrooms. Our study took place on the Eedi educational platform, an evidence-based learning ecosystem that provides students with both curriculum-aligned mathematics activities and one-to-one support from remote human tutors via online chat conversations. In our experiment, we tested whether LearnLM—a genAI model fine-tuned for pedagogical applications~\citep{jurenka2024towards, learnlm2024learnlm, learnlm2025evaluating}—could help scale this assistance.\footnote{Since the launch of this RCT, Google DeepMind has infused the pedagogical capabilities of LearnLM into the Gemini model family (see \href{https://goo.gle/learnlm}{goo.gle/learnlm} for details). Teams seeking to replicate these findings or build similar experiences should now use Gemini 2.5 Pro.} To ensure a high standard of safety and pedagogy for all students in our trial, $N = 17$ expert human tutors directly supervised LearnLM, assuming ultimate responsibility for every interaction it had with students. In particular, the tutors appraised each message that LearnLM generated, retaining full control to approve, edit, or replace it before it reached the student.

LearnLM proved to be a trustworthy source of pedagogical instruction, with the supervising tutors approving over 76\% of its messages without changes or with only minimal edits (changing one or two characters; e.g., deleting an emoji).
In fact, across all of the learning outcomes we measured, supervised support from LearnLM proved at least as effective as guidance from a human tutor. Most surprisingly, students tutored by LearnLM exhibited measurably better knowledge transfer than those receiving support from human tutors alone. On average, supervised support from LearnLM improved the probability of a student solving a novel problem correctly by 5.5 percentage points over guidance from a human tutor.

To better understand this broad effectiveness, we surveyed and interviewed the supervising tutors for their perspectives on LearnLM. They reported that LearnLM consistently generated high-quality, Socratic dialogue, providing a strong foundation for academic interactions with students. The supervising tutors' interventions tended to focus on moderating the dialogue's pacing and providing the social and emotional nuance required to maintain student engagement.

\begin{figure*}[t]
    \hspace{1em}
    \centering
    \begin{subfigure}[t]{0.55\linewidth}
        \caption{}
        \centering
        \includegraphics[width=\linewidth]{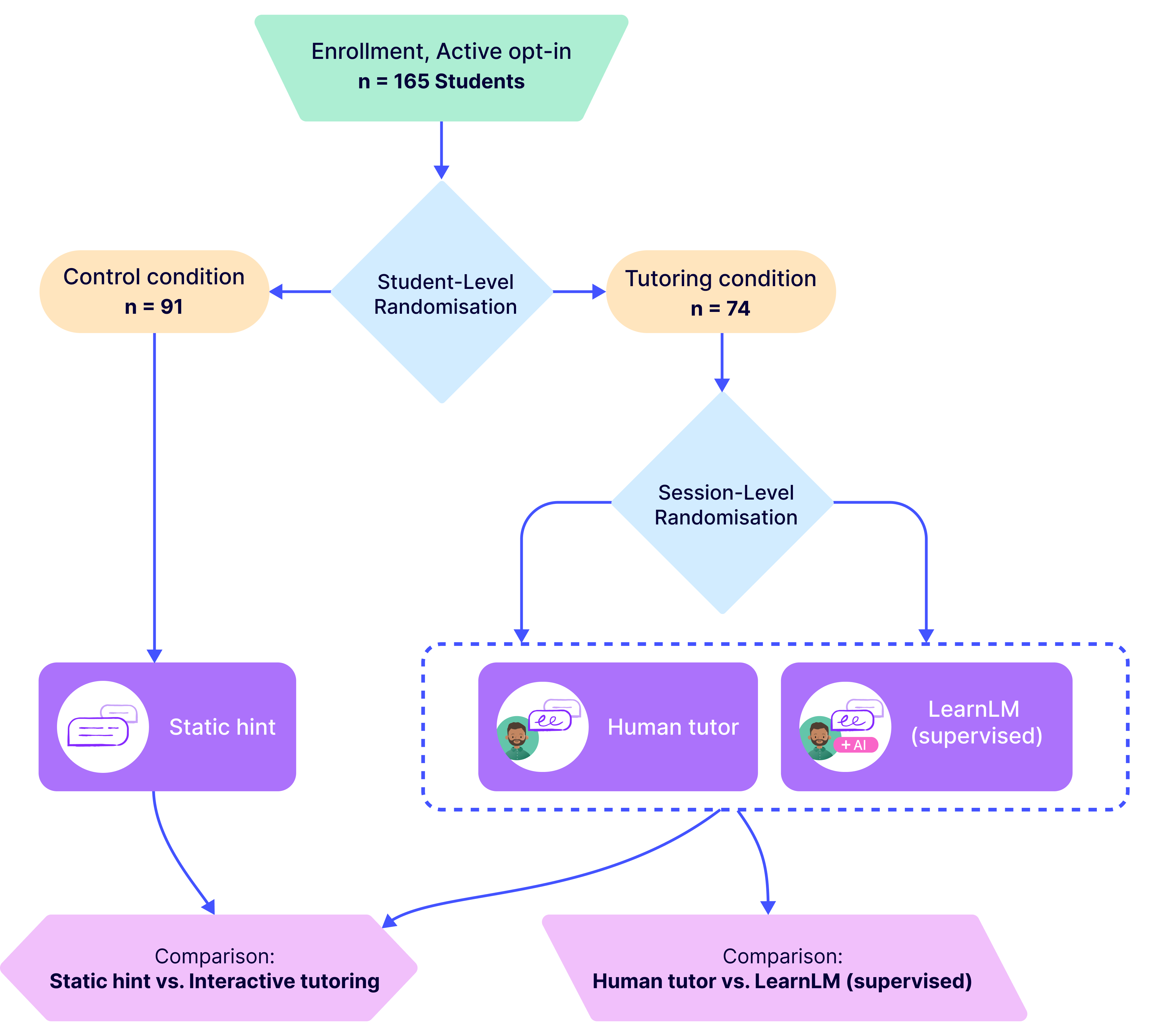}
        \label{fig:experiment_overview_a}
    \end{subfigure}
    \hfill
    \begin{subfigure}[t]{0.395\linewidth}
        \caption{}
        \centering
        \includegraphics[width=\linewidth]{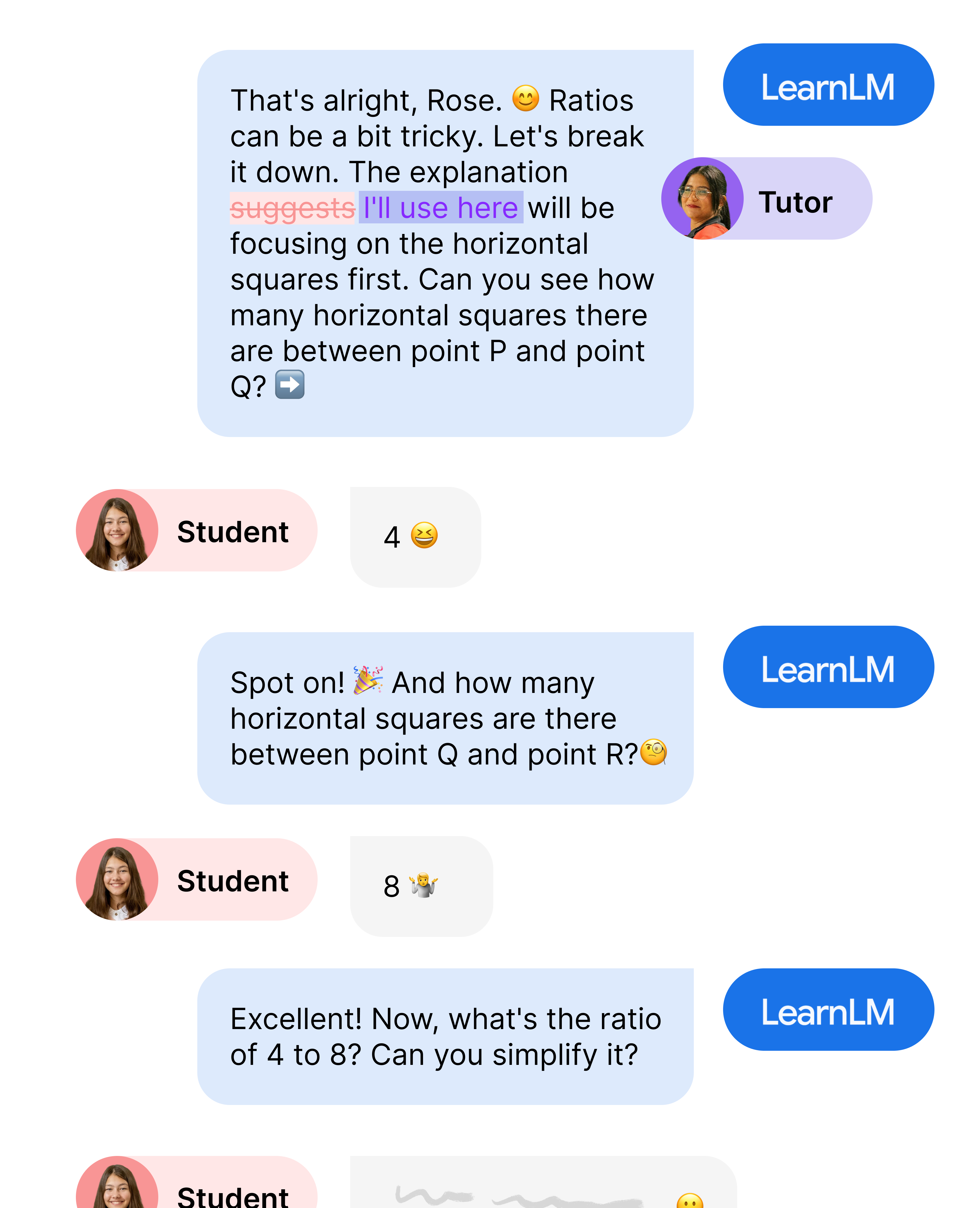}
        \label{fig:experiment_overview_b}
    \end{subfigure}
    \hspace{1em}
    \caption{We designed this exploratory RCT to evaluate the safety, pedagogy, and efficacy of LearnLM. (a) The RCT randomly assigned each of $N = 165$ students to receive either static hints or interactive tutoring. Students in the tutoring condition experienced a further level of randomization. When they started a tutoring session, the platform randomly assigned them to either a session with a human tutor or a session with LearnLM (supervised by a human tutor). This design allows us to compare static, pre-written support against interactive tutoring, as well as human tutoring against (supervised) tutoring from LearnLM. (b) In sessions with LearnLM, a supervising tutor reviewed each message that LearnLM drafted. They could either edit the message, completely re-write it, or approve it without any changes. The Eedi platform then sent the message to the student.}
    \label{fig:experiment_overview}
\end{figure*}

Overall, our exploratory RCT identifies several avenues for new research on AI and education, while also suggesting a potential role for genAI tutors in delivering effective, individualized learning support at scale.

\section{An Exploratory Classroom Trial}

Our RCT aimed to evaluate LearnLM in a rigorous, real-world, in-classroom testbed. Hundreds of secondary schools in the UK integrate the Eedi learning platform directly into their mathematics instruction. The platform provides students with curriculum-aligned study units and a spectrum of personalized support, including two forms of assistance central to this RCT: carefully designed hints for common misconceptions in each study unit, and one-to-one guidance from trained, expert tutors via online chat interactions. Students who receive this standard support on the Eedi platform experience the equivalent of two additional months of academic progress, with the impact doubling for highly engaged students~\citep{harrison2025eedi}. We recruited $N = 165$ students in Year 9 and 10 (ages 13--15) across five of these schools for the RCT  (see Appendix~\ref{sec:appendix/participants}). Each student and each tutor provided informed consent to participate in the trial. The trial ran from May through June 2025.

The trial leveraged these two forms of Eedi support—hints and chat-based tutoring (``hybrid tutoring''~\citep{chen2025vtutor})—as baselines to assess the pedagogical efficacy of LearnLM (see Figure~\ref{fig:interventions} in Appendix~\ref{sec:appendix/trial}). During the trial period, we randomly assigned each student either to receive \textit{static} pedagogical support (pre-written hints) or to enter an \textit{interactive} one-to-one tutoring session (Figure~\ref{fig:experiment_overview}; see also Appendices~\ref{sec:appendix/trial} \& \ref{sec:appendix/platform}). Students in the tutoring condition experienced a further level of randomization: when a student entered a tutoring session, we randomly connected them either with an expert human tutor or with LearnLM (supervised by a human tutor). We prompted LearnLM to adopt a Socratic approach aimed at guiding the student to identify their own mistake, and provided the model access to the full question text, the student's incorrect answer, and explanations for both the student and a teacher about the misconception underlying the incorrect answer, among other information (see Appendix~\ref{sec:appendix/model/prompt}).

Our approach allowed us to pose a set of four research questions:

\begin{enumerate}[label={\bfseries RQ\arabic*:}, nosep]
    \item Was LearnLM a reliable and pedagogically sound source of instruction?
    \item Was interactive tutoring (whether delivered by a human tutor alone or in a supervised session with LearnLM) more effective for student learning than static pedagogical support?
    \item For students receiving interactive tutoring, was support from a supervised session with LearnLM more effective than support from a human tutor working alone?
    \item What can we learn from tutor and student experiences of interacting with LearnLM?
\end{enumerate}

To answer these questions, we adopted a Bayesian framework and directly estimated the magnitude and credibility of our treatment effects. Unlike standard frequentist approaches, this method allows us to calculate the probability that one intervention outperforms another by a specific magnitude, providing a more practical foundation for making decisions about real-world deployment.
For all analyses, we assigned identical, weakly informative priors to each intervention. We then used the resulting posterior distributions to calculate the exact probability that outcomes in one group exceeded those in another, providing a more precise signal than a simple comparison of the intervals (cf.~\citep{cumming2009inference, austin2002brief, schenker2001judging}). For complete experimental details, see \hyperref[sec:methods]{\textit{Methods}} and Appendices~\ref{sec:appendix/participants}--\ref{sec:appendix/model}.

\section{Results} \label{sec:results}

We first verified the basic safety and quality of LearnLM's tutoring (\textbf{RQ1}) by auditing the full corpus of 3,617 messages that it drafted, as well as the supervising tutors' decisions to approve, edit, or rewrite those messages. LearnLM proved a trustworthy source of instruction. The tutors who supervised and reviewed its messages accepted 74.4\% without any edits.
As judged by edit distance~\citep{levenshtein1966binary, navarro2001guided}, many of the $k = 926$ instances where tutors edited or rewrote a suggestion reflected minor or targeted adjustments (see Table~\ref{tab:minor_edits} for examples). The two most frequent edit distances, accounting for 5.5\% and 2.4\% of re-writes, were just a single character and two characters, respectively; these virtually always reflected a tutor deleting or changing an emoji. The median intervention altered 59 characters, or just a few words. Still, after the RCT finished, we asked the supervising tutors to systematically review the corpus of edits and re-writes. This review revealed zero instances of harmful or risky content and only five factual errors, or 0.1\% of the total 3,617 messages that LearnLM drafted (see Table~\ref{tab:factual_errors} in Appendix~\ref{sec:appendix/edits}). Overall, a close audit confirmed that LearnLM provided safe and reliable guidance during the trial.

\begin{figure*}[t]
    \centering
    \includegraphics[width=0.95\linewidth]{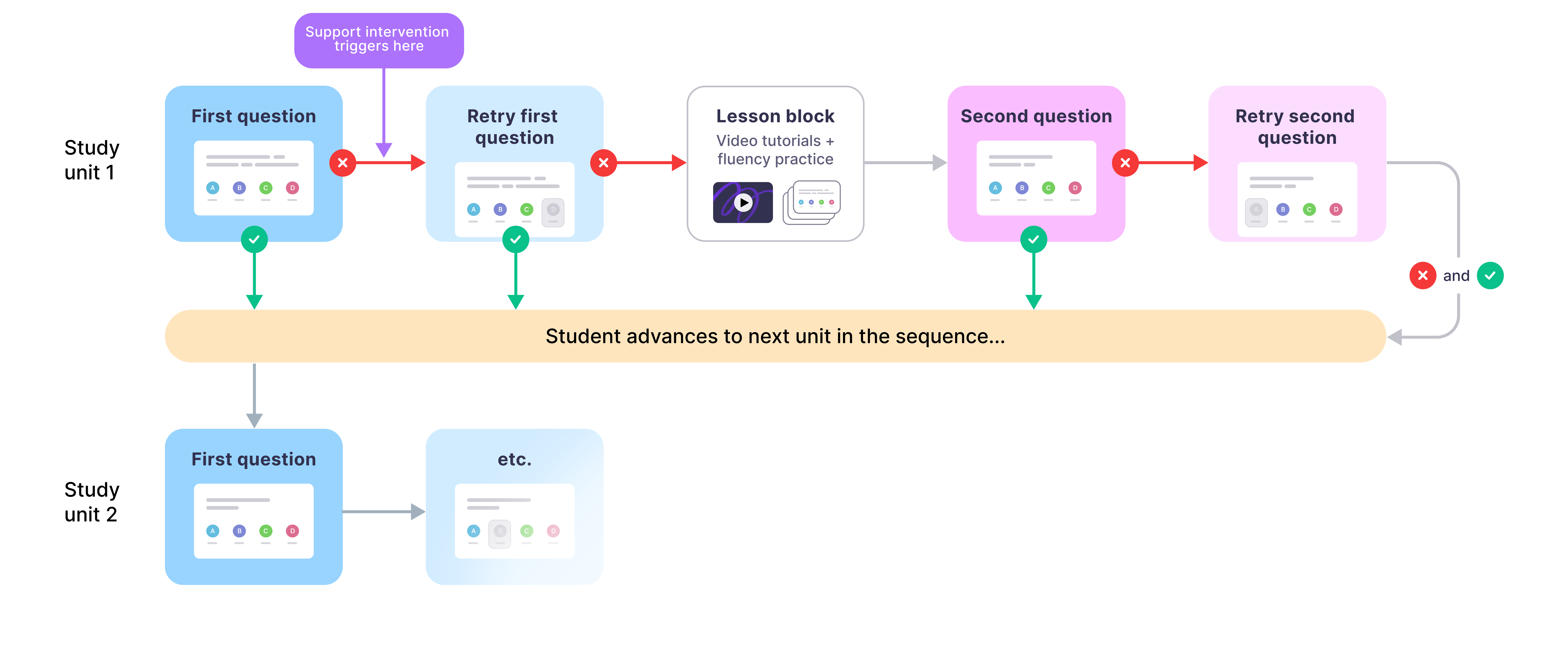}
    \caption{Student progression through the study unit. If a student makes a mistake on the first question in a study unit, they receive a support intervention. We analyze whether the intervention helps the student identify and remediate their mistake, resolve the misconception underlying their incorrect choice, and transfer the knowledge from the intervention to the next study unit. See \hyperref[sec:methods]{\textit{Methods}} and Appendix~\ref{sec:appendix/platform} for more information on the Eedi platform.}
    \label{fig:study_unit}
\end{figure*}

Next, we evaluated effects on student learning (\textbf{RQ2}, \textbf{RQ3}), comparing students' performance after receiving one of the standard interventions on the Eedi platform or interacting with LearnLM. As described in \hyperref[sec:methods]{\textit{Methods}}, students worked through a series of short study units, each consisting of several multiple-choice questions designed to assess a specific mathematical topic (Figure~\ref{fig:study_unit}). Whenever a student answered the first question in a unit incorrectly, the platform triggered a support intervention. Depending on their assigned condition, students either received a static, pre-written hint specific to their mistake on that question, or an interactive (chat-based) session with a tutor. Immediately following the intervention, the platform presented the student with the exact same question and prompted them to try answering it again.

\begin{figure*}[t]
    \centering
    \includegraphics[width=\linewidth]{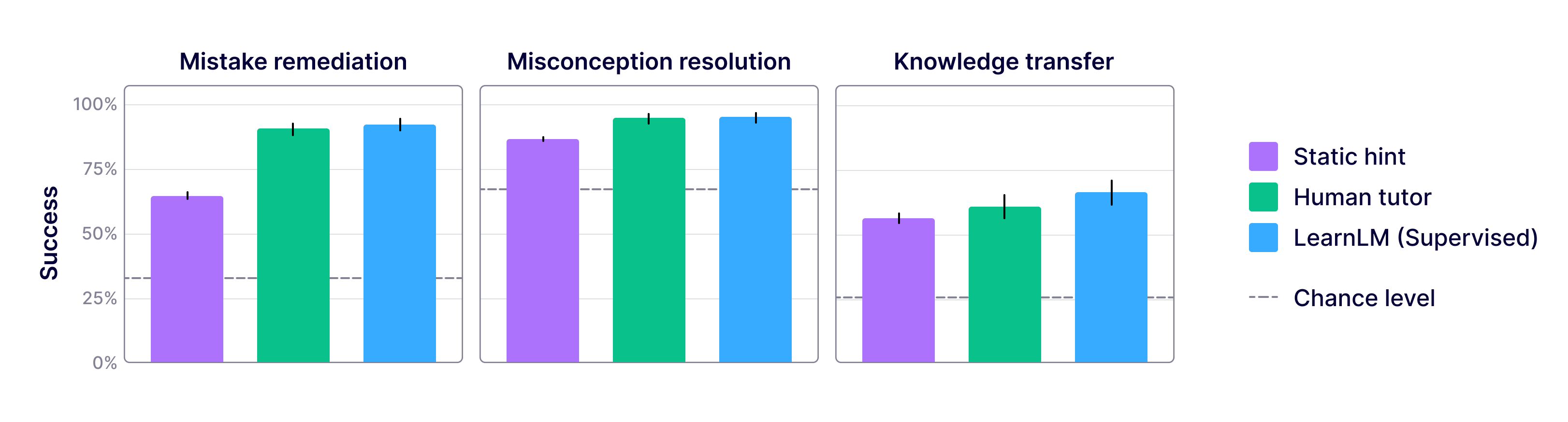}
    \caption{Tutoring interventions improve student learning outcomes. (left, center) For immediate learning outcomes, sessions with human tutors and expert-supervised sessions with LearnLM promote similar growth for students. Students who receive interactive tutoring from either source substantially outperform students who receive pre-written, static hints. (right) In contrast, students tutored by LearnLM demonstrate greater knowledge transfer to new topics than those supported either by static hints or by human tutors alone. Error bars indicate 95\% credible intervals. Dashed lines represent the chance of success when guessing randomly (33.3\%, 66.7\%, and 25\%, respectively).}
    \label{fig:learning_outcomes}
\end{figure*}

Echoing prior research~\citep{nickow2020impressive}, interactive support with a human tutor proved far more effective for this kind of immediate course-correction. Students who joined a real-time session with a human tutor were substantially more likely to correct their mistakes than were those who received a static, pre-written hint (see Figure~\ref{fig:learning_outcomes}, left). In particular, 91.2\% of students who received interactive support from a human tutor solved the problem correctly on their second attempt (with a 95\% credible interval of [88.5\%, 93.6\%]), compared to only 65.4\% [63.8\%, 66.9\%] of students who received a static hint. Supervised instruction from LearnLM proved just as effective at helping students correct their mistakes. Students receiving guidance from LearnLM answered their second attempt correctly 93.0\% [90.4\%, 95.3\%] of the time. (For context, simply eliminating the previous mistake and guessing from the remaining options would yield an expected success rate of 33.3\%.)

If a student still answered the question incorrectly on their second attempt, the platform provided them with several additional opportunities to correct their underlying misconception. Specifically, it offered them two attempts at a new question on the exact same mathematical topic. We thus examined whether tutoring helped students eventually resolve their misunderstanding—that is, whether they answered \textit{any} of the post-intervention questions correctly. On this broader measure, interactive tutoring once again proved superior to static hints  (see Figure~\ref{fig:learning_outcomes}, center). When working with a human tutor, 94.9\% [92.6\%, 96.8\%] of students resolved their misconception, relative to only 86.8\% [85.7\%, 88.0\%] of students receiving pre-written hints. No meaningful difference emerged between students working with LearnLM and those working with human tutors. Students tutored by LearnLM resolved misconceptions 95.4\% [93.1\%, 97.1\%] of the time. For this kind of near-term correction, both interactive methods appear equally effective.

Of course, the critical question is whether these guided successes (the opportunity to remediate mistakes and resolve misconceptions) reflect durable learning (the ability to solve a new problem without any assistance). Within the scope of this RCT, the best test for durable effects of tutoring is how students performed when progressing to a new topic. The Eedi platform organizes study units into sequences of five, where each unit builds directly upon the last. Our subsequent analysis therefore analyzed a student's likelihood of correctly answering the initial question in the very next unit in their current sequence.

Here, a clear advantage for LearnLM's tutoring emerged (see Figure~\ref{fig:learning_outcomes}, right). Students tutored by LearnLM on a study unit proved substantially more likely to answer the first question in the following unit correctly (66.2\% [61.1\%, 71.2\%]) than students who had received help from an unassisted human tutor (60.7\% [55.8\%, 65.4\%]). In particular, a supervised session with LearnLM increased the likelihood of learning transfer to a distinct topic by an additional 5.5 percentage points [--1.4\%, +12.4\%] relative to human tutoring. Both tutee groups, in turn, outperformed students who had received only a static hint (56.2\% [54.2\%, 58.2\%]). Altogether, we attribute a high credibility (93.6\%) to LearnLM offering better support for knowledge transfer than human tutors alone, and near certainty (>99.9\%) to its advantage over static, pre-written hints. The AI-supported interventions fostered a more durable and transferable understanding—an advantage revealed only when students faced a fresh challenge. (See Appendix~\ref{sec:appendix/learning} for our full analysis of learning outcome data.)

Throughout the RCT, we sought a richer, more nuanced understanding of the experience of interactions with LearnLM (\textbf{RQ4}) by conducting in-depth, semi-structured interviews with a random subset of $N = 5$ supervising tutors (see Table~\ref{tab:interviewed_tutors}). In addition, we invited all students and supervising tutors to share their thoughts in brief surveys. We gathered $N = 27$ student responses from a post-trial survey, and $N = 17$ tutor responses on both pre- and post-trial surveys. These firsthand perspectives help contextualize LearnLM's effectiveness and the specific role that human expertise played in its tutoring successes.

\begin{table}[b]
    \centering
    \begin{tabular}{@{}lll@{}}
    \toprule
    \textbf{Tutor ID} & \textbf{Gender} & \textbf{Years of teaching experience} \\
    \midrule
    T1 & F & 6--10 years \\
    T2 & F & More than 10 years \\
    T3 & F & More than 10 years \\
    T4 & F & More than 10 years \\
    T5 & F & 6--10 years \\
    \bottomrule
    \end{tabular}
    \caption{We conducted semi-structured interviews with a subset of five supervising tutors to seek a deeper, nuanced understanding of LearnLM's behavior and the general experience of participating in the RCT. Table~\ref{tab:all_tutors} in Appendix~\ref{sec:appendix/participants} contains comparable details for the full sample of supervising tutors.} \label{tab:interviewed_tutors}
\end{table}

Over the course of the trial, supervising tutors came to view LearnLM as a source of high-quality, expert-level pedagogical insights. The most prominent theme from our interviews, raised independently by all five interviewed tutors, was LearnLM's consistent use of Socratic dialogue. Tutors reported that its suggestions prompted a more inquisitive, student-led interaction. One tutor highlighted its ability to ask ``really good questions that I hadn't necessarily thought of [...] in a good way, a nice way'' (T3). As another reported, ``[LearnLM] definitely explained certain topics in a better way than I probably could have'' (T5). This praise aligned with tutors' actions during the trial: as established earlier, the tutors approved the vast majority of the messages drafted by LearnLM without any edits or changes.

In interviews, three tutors noted that supervising this high standard of instruction prompted an unexpected outcome on their part: professional growth and development. For instance, one tutor contrasted LearnLM's Socratic strategy with their prior approach, noting that the drafted messages prompted ``questions more like `Okay, what made you think that was the answer?' [...] whereas before [...] my main goal was to identify their misconception myself'' (T1). Another explained, ``I remember thinking, `Oh, I hadn't thought of explaining it that way before.' Just like when you watch another teacher'' (T2). Over the course of the trial, LearnLM's standard of instruction made a considerable impression on the tutors who supervised it.

One-to-one tutoring requires sustained, substantive effort to process the scenario at hand and craft effective pedagogical guidance. LearnLM's ability to consistently generate high-quality pedagogical responses thus made the entire tutoring process more fluid and efficient. Our post-trial surveys corroborated this; when asked about LearnLM's most useful feature, 82.4\% of tutors chose ``supporting multiple students at the same time.'' This new, effective process quickly set a new standard for the supervising tutors. In fact, every tutor that we interviewed independently raised this increased capacity as a key strength. As one tutor explained in their interview, ``I got to the point of being disappointed when I didn't get [a session] with the AI suggestions'' (T2). These positive experiences translated into a broad increase in comfort with AI across the cohort. Tutors' self-reported comfort with using AI tools rose from an average of 3.4 [2.9, 4.0] out of 5 in the pre-trial survey to 3.9 [3.3, 4.4] in the post-trial survey (posterior probability of increase: 90.0\%).

Building LearnLM's pedagogical insights into effective tutoring conversations, however, required the supervising tutors to incorporate social and emotional nuance from their understanding of the students. Our retrospective analysis of the 25.6\% of cases where tutors edited or re-wrote LearnLM's messages identified two primary motivations for these interventions: moderating the pedagogical pacing of the conversation and providing social-emotional nuance to LearnLM drafts. The most frequent intervention was adjusting the conversation's pacing to prevent exasperating students, accounting for 44.3\% of all edits. Our tutors echoed this specific challenge in five of our five interviews. As one tutor explained in their interview, ``quite often the students just got frustrated, and then they lost complete interest in the question, so it was a case of overriding it'' (T2). Tutors often found it necessary to step in when LearnLM's Socratic questions, while pedagogically sound, persisted longer than a student's patience. One tutor described a common scenario where ``[LearnLM] will go, `Okay, you've got the answer. Let's dig a little deeper about why you've got that answer.' And the child is just like, `No, I've got it. I know what I'm doing. Can I go now?''' (T1).

Providing social and emotional context to LearnLM's drafts emerged as a second prominent motivation for supervisors' interventions. In total, 19.5\% of tutors' edits adjusted the persona or tone conveyed by the drafted messages. Tutors consistently added personal touches that recognized the student as an individual. For example, one tutor noted the importance of acknowledging a student they had helped before, a nuance LearnLM could not replicate, given that its prompt did not provide any information on past tutoring sessions: ``...if you'd already helped that student twice before, [LearnLM] didn't quite have the capability to go like, `Oh Sarah, it's you again. Hi!' And I like to have that kind of rapport'' (T3). Tutors also calibrated the tone of the messages to ensure they were appropriate for student communication styles. One tutor remarked that LearnLM's predilection for emojis ``comes across as a bit fake, and [...] the students pick up on that'' (T1). Overall, the human tutors grounded LearnLM's suggestions with social and emotional nuance, translating its pedagogical insights into effective educational interactions.

Finally, student feedback indicated broad satisfaction with their tutoring interactions. In post-trial surveys, students who received interactive tutoring rated the helpfulness of the support they received an average of 3.9 [3.1, 4.7] out of 5, relative to 3.6 [2.9, 4.2] for students who received static hints (posterior probability of an advantage for tutoring: 74.9\%). Ultimately, interactive tutoring delivered not just strong learning outcomes, but an enjoyable experience for the learners themselves.

\section{Discussion}

When deployed responsibly, can generative AI safely and effectively support students in real-world learning environments? Our exploratory trial investigated whether LearnLM—a genAI model fine-tuned for pedagogical applications—could help provide in-classroom guidance across five UK secondary schools. Students in these schools use Eedi, an online mathematics platform that effectively improves learning outcomes~\citep{harrison2025eedi}, for their regular instruction. We incorporated LearnLM into the platform
so that it drafted messages to send to students in chat-based tutoring sessions. Of course, genAI tools carry well-known risks, including their capacity to fabricate information~\citep{sallay2024vetting, maylahn2024state} and erode critical thinking~\citep{gerlich2025ai, laird2025hand}. Given the heightened ethical weight of these risks in educational settings, we assigned a group of expert (human) tutors to directly supervise LearnLM, assuming ultimate responsibility for each of its interactions with students.
The tutors applied a simple, rigorous standard: they revised each of LearnLM's drafts until they were satisfied sending the message as their own.

The supervising tutors found LearnLM to be a reliable source of pedagogical instruction, approving the vast majority of its drafted messages without any edits. A systematic review of the drafted messages revealed zero instances of harmful content and only five factual errors out of 3,617 messages drafted by LearnLM total. For students, this translated into effective support for learning: tutoring from LearnLM helped students identify their mistakes and correct their misconceptions just as well as instruction from human tutors alone. Unexpectedly, students tutored by LearnLM demonstrated greater knowledge transfer to subsequent topics than did students who received guidance from human tutors.

Tutors consistently praised LearnLM's use of Socratic dialogue, but also noted that the model's relatively inflexible adherence to pedagogical principles threatened to exasperate some students. The best human tutors, in contrast, draw on experience, empathy, and judgment to decide when to push students and when to moderate their approach. This is a constant judgment call for tutors: weighing the long-term benefits of productive struggle against the immediate risks of frustrating a student and causing them to disengage completely. This delicate calibration remains a fundamental challenge for current AI systems~\citep{dibenedetto2025gpt, dibenedetto2025gemini, dibenedetto2025claude, gillick2025ai}.

Beyond safety and pedagogy, expanding access to one-to-one tutoring will require improving its cost and scalability. In our interviews, the supervising tutors consistently reported that LearnLM made their work feel more fluid and efficient. Our own anecdotal observations during the trial supported these reports: tutors appeared comfortable managing higher workloads during their supervised sessions. Unfortunately, the design of this RCT—with tutors fluidly switching from supervision to direct interaction during the same classroom periods—precludes a rigorous measurement of throughput or efficiency for each condition. After the trial, we simulated additional sessions as an informal test of scalability (see Appendix~\ref{sec:appendix/operational}). The results of this informal test corroborate the improved efficiency of the supervised sessions, with tutors sustaining a higher volume of simultaneous conversations when supported by LearnLM. Altogether, these signals support a possible role for genAI tutoring in helping educators to deliver individualized instruction at scale.

Overall, the design of this exploratory RCT allowed us to rapidly validate LearnLM's safety and gather initial signals of its efficacy. We measured these outcomes using students' standard, daily activities on the Eedi platform. This approach provided us with learning signals immediately, eliminating the need to develop and administer new trial-specific assessments, or to wait for the next round of standardized exams. In addition, by randomly assigning the source of support for each individual tutoring session, we could measure the alternating impact of LearnLM and human tutoring on the same students. This approach disentangled tutoring effectiveness from pre-existing student differences, permitting us to detect meaningful indications of efficacy working with just five schools.

On the other hand, this design offers only a partial glimpse at the broader trajectory of learning. Randomizing the source of tutoring session-by-session allowed our RCT to efficiently investigate immediate learning outcomes, but also prevented it from isolating the cumulative impact of working with LearnLM over time. Measuring substantive, longer-term effects on learning will require a different approach. In addition, the finding from our interviews that tutors learned from supervising LearnLM indicates another methodological wrinkle. If tutors applied those insights in sessions without LearnLM, that crossover might dampen the measured difference between the two tutoring conditions. Future research can overcome these limitations by assigning students to receive one consistent type of support for an entire study, ideally following their progress over several months and tracking their performance on external, standardized assessments. Such a longitudinal approach could help determine whether the immediate successes that we observed translate into persistent, substantive learning gains—a vital step toward validating the potential of AI tutoring to deliver scalable, individualized support for students and educators.

To what extent might the tutoring efficacy we observe in this RCT generalize beyond mathematics? In part, LearnLM's strong performance reflects the nature of the inputs that we provided to it: questions with precise answers, discrete incorrect responses, and validated explanations of why students might have veered off the right path. Mathematics curricula often focus on verifiably solvable problems, so they readily offer this clear structure. In contrast, many other subjects taught in secondary school emphasize ambiguity, interpretation, and argumentation. Consequently, LearnLM's performance in this trial offers limited evidence for its ability to shepherd students through more interpretive activities in fields like history or literature. We will need to conduct research across a diverse range of subjects to understand where current AI tutors may already offer strong support, and which domains require us to develop new, distinct approaches to AI pedagogy.

Ultimately, our research did not start from scratch with this trial. Two lines of conceptual and empirical groundwork enabled this RCT: first, a generative AI model specifically fine-tuned for pedagogy~\citep{jurenka2024towards, learnlm2024learnlm, learnlm2025evaluating}, and second, an educational platform deeply rooted in learning science~\citep{harrison2025eedi}. Our results integrating LearnLM into the Eedi ecosystem illustrate how learning science and technological development can complement one another to support and scale better learning outcomes for students. Moving forward, we invite collaboration across the AI and learning science communities to partner on new research and offer an honest appraisal of how this technology helps—or hurts—students and educators in different contexts and settings. Building and sharing this knowledge helps bring us closer to the goal of providing effective, safe, and accessible learning opportunities for all students.

\section{Methods} \label{sec:methods}

Our protocol underwent independent ethical review, with a favourable opinion from the Human Behavioural Research Ethics Committee at Google DeepMind (\#25 003).

\paragraph{Participants} We recruited $N = 165$ students from five UK secondary schools to participate in the trial. We drew the cohort exclusively in Years 9 and 10 (ages 13--15), from classrooms that incorporate the Eedi platform as part of their regular mathematics instruction for one hour per week. Each student provided informed consent to participate in this research. As part of their informed consent process, we explained to students that their tutors might rely on AI support during the trial. A pool of $N = 17$ expert tutors—all qualified teachers with extensive teaching experience—delivered the trial's interactive interventions (i.e., tutored students directly and supervised tutoring sessions with LearnLM). Each tutor also provided informed consent to participate in this research.

\paragraph{Platform} The Eedi platform provides a range of curriculum-aligned mathematics activities for students and classrooms. In this RCT, we focused on student performance on its short study units, each designed to assess a specific mathematics topic and consisting of diagnostic multiple-choice questions with four response options (Figure~\ref{fig:study_unit}). Whenever a student answers the first question in a unit incorrectly, the platform triggers a support intervention. Immediately following this intervention, the platform prompts the student to retry the question that they originally missed. If they miss this question again, the platform presents them with a new question on the same topic, written to assess the same topic and misconceptions using different concrete details. Students complete a unit and progress to the next unit as soon as they answer a question correctly, or after they incorrectly answer all four questions. The platform organizes these study units into sequences of five. Individual study units in a sequence build iteratively upon one another, so students must typically grasp one before successfully engaging with the next.

\paragraph{Model} LearnLM is a family of generative AI models fine-tuned to specialize in pedagogical dialogue.\footnotemark[1] For this RCT, we accessed the most recent version of LearnLM available at the time, fine-tuned from Gemini 2.0 Flash. We connected the Eedi platform to LearnLM via a custom API created specifically for this trial. During platform tutoring sessions with LearnLM, the platform assembled a strictly defined system prompt instructing the model to draft a concise, Socratic response aimed at guiding the student to self-correct their specific misconception without revealing the answer. The prompt also provided rich real-time context, including the question text, the student's incorrect answer, and the specific misconception underlying the answer identified by the platform (see Appendix~\ref{sec:appendix/model/prompt} for the detailed prompt). The platform sent the assembled prompt to the API, which then returned a draft response from LearnLM for the platform to pass to the supervising tutor for approval, editing, or re-writing.

\paragraph{Procedure} We conducted the exploratory RCT over seven consecutive weeks (May through June 2025). The trial employed a two-level randomized controlled design to address our research questions. First, we randomly assigned students to either the control condition ($N = 91$ students) or the tutoring condition ($N = 74$ students). Second, specifically for students in the tutoring condition, we randomly assigned each individual tutoring session to either a human expert or to LearnLM (under supervision from a human expert).

Whenever a student in the control condition answered a question incorrectly, they received a pre-written message designed to prompt reflection on a specific misconception, based on which incorrect option they selected (a ``static hint''). The platform then prompted them to retry the question.

To support the tutoring condition, we scheduled a team of tutors to remain on-call in the Eedi platform during class hours on each day of the trial. Whenever a student in the tutoring condition answered a question incorrectly, the standby team received an alert. One of the tutors would then initiate a session with the student. The platform randomized each of these sessions to either connect the tutor directly to the student (``session with a human tutor alone'') or to assign them to supervise the model (``supervised session with LearnLM''). That is, tutors both directly guided students and oversaw sessions with LearnLM on the same day. In supervised sessions with LearnLM, the human tutor reviewed the suggestions generated by the model and approved, edited, or replaced each drafted message before the platform sent it to the student. The student interface appeared identical across both conditions, with no explicit indication of whether the student was connected with a human tutor alone or a tutor supervising LearnLM.

For both conditions, we recorded the student and question identifiers, timing, correctness, and position (both within its study unit and within its sequence of five units) of every attempted answer on the platform.

To complement this central evaluation, we incorporated several qualitative lines of inquiry. First, we recorded the entire message corpus and the supervising tutors' decisions. Throughout the seven-week trial, the platform logged every draft generated by LearnLM, the supervising tutor's action (approve, edit, or re-write), and the finalized message sent to the student. Second, we administered short baseline and endline surveys to all supervising tutors. All tutors completed both rounds ($N = 17$). Third, we invited all participating students to complete a short survey via the Eedi platform after the trial concluded, resulting in $N = 27$ responses. Finally, we randomly selected five tutors and invited them to participate in hour-long, semi-structured interviews. These interviews followed a standardized protocol designed to elicit detailed narratives of their experiences supervising LearnLM.

\paragraph{Analysis.} We evaluated efficacy across three primary quantitative outcomes derived from Eedi platform data: mistake remediation (success at attempting a question a second time, after an intervention), misconception resolution (success at answering any question within a study unit, after an intervention), and knowledge transfer (success at answering the first question of the next study unit  within the same sequence, after an intervention). We leveraged Bayesian regression to estimate treatment effects for these outcomes. We included baseline performance as a covariate in all regression models to account for pre-existing differences between students. The success rates reported in the \hyperref[sec:results]{\textit{Results}} section represent posterior predictive margins estimated from these regressions, adjusting for students' baseline performance. Practically speaking, these estimates differ only negligibly from the unadjusted success rates observed during the trial (see Appendix~\ref{sec:appendix/learning} for all unadjusted success rates and posterior predictive margins).

To verify the safety and pedagogical quality of LearnLM's tutoring, we audited the full corpus of drafted messages through an iterative, inductive process~\citep{braun2006using}. We first counted the number of outright approvals without changes. For all edited and re-written messages, we quantified the magnitude of change by computing the Levenshtein distance and the edit ratio (the Levenshtein distance divided by the total character count of the initial draft). We then categorized the apparent functional purpose of each revision. Specifically, a generative AI model (Gemini 2.5 Pro~\citep{gemini2025gemini}) performed an initial coding of every edit, processing 30 to 50 pairs of original and edited messages at a time. Two members of the research team reviewed and refined the generated codes into a focused codebook. Next, two expert tutors reviewed each pair of messages to validate the assigned codes. A member of the research team then conducted a final review of the coding decisions to ensure consistency and accuracy. Finally, the research team synthesized these codes into broader themes and specifically searched the coded corpus for any instances of harmful or erroneous generations.

We took an iterative approach to identify themes in the supervising tutors' interviews, following emerging guidance on applying genAI tools to support qualitative coding~\citep{bastani2025generative, hitch2024artificial, nyaaba2025optimizing}. A member of the research team first reviewed all transcripts to gain familiarity with the content. We then applied a generative AI model (Gemini 2.5 Pro) to identify segments of text describing tutors' perceptions, experiences, or attitudes and to generate initial descriptive labels for them.
A member of the research team then refined them into clear definitions, organized them into a structured set of themes, and then manually applied these labels to the full dataset. Finally, a member of the research team verified every coded excerpt against the original transcript to create a complete audit trail.

Finally, we analyzed responses from our short surveys for additional context on student and tutor experiences and perspectives.

\color{black}

\bibliography{main}

\clearpage
\section*{Contributions and Acknowledgments}

\subsection*{Core Contributors}

The following individuals contributed to the work described in this report. These lists are ordered alphabetically, and do not indicate ranking of contributions.

On the Google team, the following individuals made core contributions:

\noindent
Albert Wang,
Aliya Rysbek,
Andrea Huber,
Brian Veprek,
Irina Jurenka,
Jonathan Caton,
Julia Wilkowski,
Kaiz Alarakyia,
Kevin R. McKee,
Liam McCafferty,
Markus Kunesch,
Sara Wiltberger,
and
Shakir Mohamed.

On the Eedi team, the following individuals made core contributions:

\noindent
Anna Kenolty,
Anjali Nambiar,
Ben Caulfield,
Beth Lilley-Draper,
Bibi Groot,
Chelsea Burdett,
Claire Willis,
Craig Barton,
Digory Smith,
George Mu,
Harriet Walters,
Iris Hulls,
James Stalley-Moores,
Lucy Dalton,
Pauline Malubay,
Rachel Kidson,
Rich Wells,
Sam Wheeler,
Simon Woodhead,
and
Vasco Brazão.

Kevin R. McKee and Bibi Groot led this research and the preparation of this report.

\subsection*{Acknowledgements}

This work represents a close collaboration between Google and Eedi.

For Eedi: We would like to acknowledge the support of the Eedi tutors and learning designers who made this work possible, including
Bea Pugh,
Gemma Bazany-Barber,
Marion Brehm,
Morgan Sowerby,
Nigel Kendall,
and
Zoe Sutcliffe.
Eedi also thanks the organizations that fund our research efforts: the Digital Harbor Foundation, Learning Engineering Virtual Institute, Rockefeller Foundation, Schmidt Futures, and Walton Family Foundation.

For Google: We completed this work as part of the LearnLM effort---a cross-Google project, with members from Google DeepMind, Google Research, Google LearnX, and more.
This tech report represents only a small part of the wider effort, and only lists team members who made direct contributions to this report.
The dedication and efforts of numerous teams make our work possible. The LearnLM team would like to acknowledge support from
Abhinit Modi,
Aditya Srikanth Veerubhotla,
Antonia Mould,
Avishkar Bhoopchand,
Brett Wiltshire,
Daniel Gillick,
Daniel Kasenberg,
Derek Ahmed,
Gal Elidan,
James Cohan,
Jennifer She,
Kristen Morea,
Lisa Wang,
Mike Schaekermann,
Miriam Schneider,
Miruna Pîslar,
Muktha Ananda,
Nahema Marchal,
Nikhil Joshi,
Parsa Mahmoudieh,
Paul Jhun,
Shanice Onike,
Shashank Agarwal,
Shubham Milind Phal,
Sun Jae Lee,
Theofilos Strinopoulos,
Wei-Jen Ko,
and
Will Hawkins.
Furthermore, we would like to thank the Gemini team, the Compute team, the Responsible Development and Innovation team, the Responsible Engineering team, and the Child Safety team at Google DeepMind, as well as the Trust and Safety team at Google.
Finally, the LearnLM team would like to acknowledge the support provided by our leads and sponsors that made this project possible: our genuine thanks go to Benedict Gomes, Lila Ibrahim, and Zoubin Ghahramani.

\onecolumn
\appendix

\counterwithin{figure}{section}
\counterwithin{table}{section}
\counterwithin{lstlisting}{section}

\section{Participants} \label{sec:appendix/participants}

\subsection{Students} \label{sec:appendix/participants/students}

The trial included $N = 165$ students in Year 9 and 10 (ages 13--15) from five UK secondary schools. Participants ranged in age from 13 to 15. Among those who reported their gender, the cohort was relatively evenly split (51.1\% female, 48.9\% male).

The schools varied broadly in academic performance and socio-economic background. Progress 8 scores ranged from --0.68 to +0.24, spanning the 5th to 75th national percentiles for state-funded schools in England~\citep{andrews2017introduction}. Free School Meal eligibility ranged from 12\% (representing affluent areas) to 26\% (closely aligned with the national secondary school average of 25.7\%~\citep{dfe2025schools}). However, the schools contained low proportions of students speaking English as an Additional Language (EAL), ranging from 2--11\%. These rates fall below the national average and do not reflect the EAL rates seen in major urban centers~\citep{lindorff2025english}.

\subsection{Tutors} \label{sec:appendix/participants/tutors}

A pool of $N = 17$ expert tutors delivered the interactive interventions in the RCT and provided additional insights in baseline surveys, semi-structured interviews, and endline surveys (Table~\ref{tab:all_tutors}).

\begin{table}[h]
    \centering
    \begin{tabular}{@{}llll@{}}
    \toprule
    \textbf{Tutor ID} & \textbf{Gender} & \textbf{Years of teaching experience} & \textbf{Additional contributions} \\
    \midrule
    T1 & F & 6--10 years & Interview, surveys (baseline, endline) \\
    T2 & F & More than 10 years & Interview, surveys (baseline, endline) \\
    T3 & F & More than 10 years & Interview, surveys (baseline, endline) \\
    T4 & F & More than 10 years & Interview, surveys (baseline, endline) \\
    T5 & F & 6--10 years & Interview, surveys (baseline, endline) \\
    T6 & F & More than 10 years & Surveys (baseline, endline) \\
    T7 & F & More than 10 years & Surveys (baseline, endline) \\
    T8 & F & 3--5 years & Surveys (baseline, endline) \\
    T9 & F & 6--10 years & Surveys (baseline, endline) \\
    T10 & M & 6--10 years & Surveys (baseline, endline) \\
    T11 & F & More than 10 years & Surveys (baseline, endline) \\
    T12 & F & 6--10 years & Surveys (baseline, endline) \\
    T13 & F & More than 10 years & Surveys (baseline, endline) \\
    T14 & F & More than 10 years & Surveys (baseline, endline) \\
    T15 & F & More than 10 years & Surveys (baseline, endline) \\
    T16 & F & 3--5 years & Surveys (baseline, endline) \\
    T17 & F & More than 10 years & Surveys (baseline, endline) \\
    \bottomrule
    \end{tabular}
    \caption{Teaching experience and additional research contributions for all supervising tutors.}
    \label{tab:all_tutors}
\end{table}

\clearpage
\section{Trial}  \label{sec:appendix/trial}

\begin{figure*}[h]
    \centering
    \includegraphics[width=\linewidth]{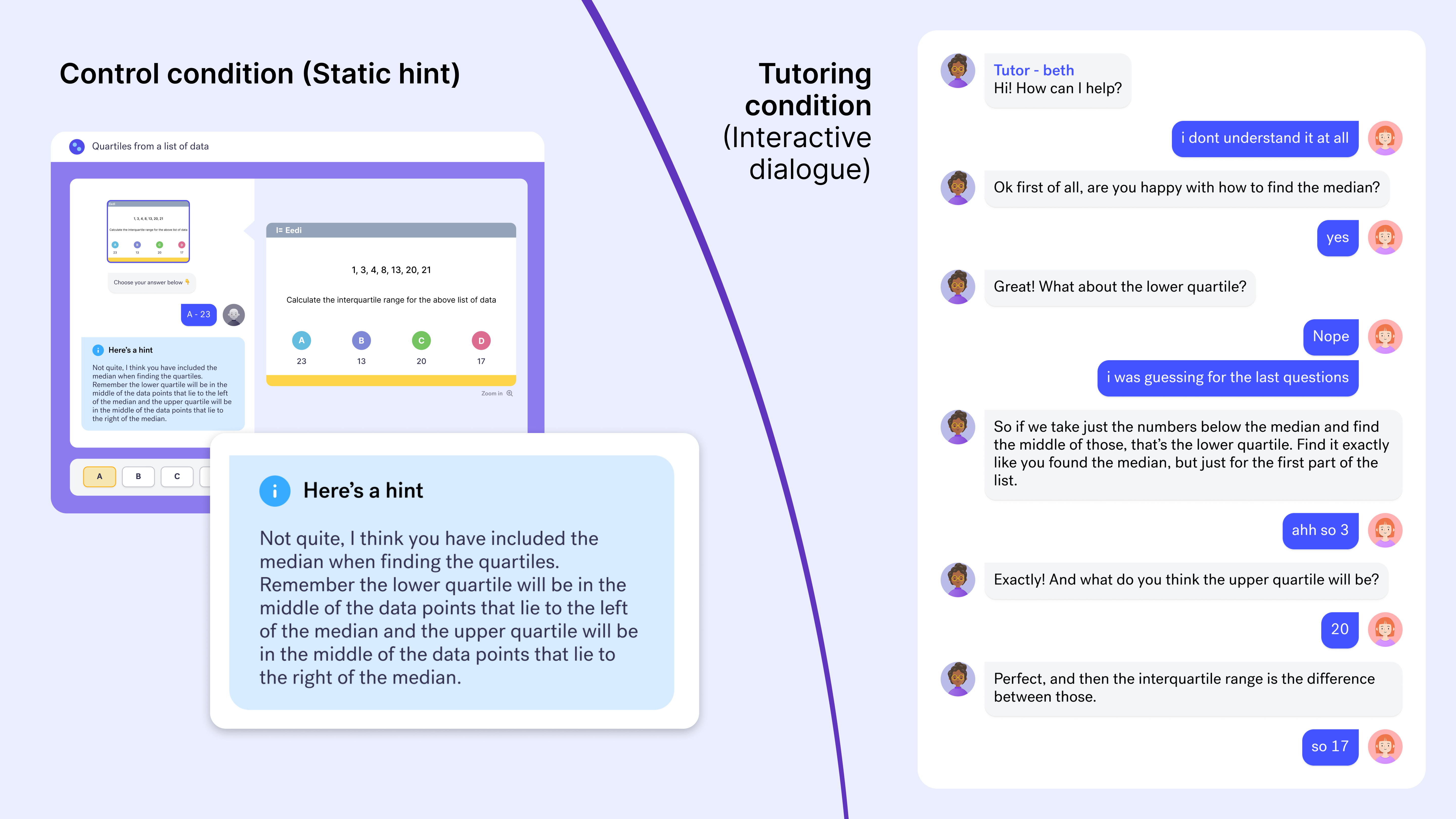}
    \caption{Our RCT focused on two support interventions on the Eedi platform. After making a mistake in a study unit, students in the control condition received static hints (left), which deliver immediate, pre-written feedback targeting the specific misconception underlying the incorrect answer they chose. Students in the tutoring condition (right) received one-to-one, chat-based assistance from a tutor.}
    \label{fig:interventions}
\end{figure*}

The research presented in this report focuses on two types of support interventions provided by the Eedi platform: static, pre-written hints that map to particular student misconceptions about individual topics, and interactive, chat-based tutoring (Figure~\ref{fig:interventions}).

To estimate baseline performance levels, we examined data from regular platform usage during the ten weeks preceding the trial, from March 1 to May 12, 2025 (the baseline period). During this period, the platform provided all students with static hints when they answered the first question of a study unit incorrectly.

Following the baseline period, we conducted the RCT over seven consecutive weeks, from May 13 to June 30, 2025 (the trial period). At the start of the trial, we randomly assigned each student to one of two conditions. Students in the control condition continued to receive only static hints after they made a mistake on the initial question of a study unit. Whenever a student in the tutoring condition answered the first question of a study unit incorrectly, the platform instead initiated an interactive, chat-based tutoring session for them. The students in the tutoring condition experienced a further level of randomization: at the start of each of their tutoring sessions, the platform randomly connected the student either to a human tutor working alone or to a supervised session with LearnLM.

Because the Eedi platform dynamically triggered support interventions based on students' real-time performance, the trial did not follow a fixed schedule. Beyond these platform-initiated support interventions, students in both conditions could also manually request tutoring support at any time. In addition, the platform allowed students to cancel tutoring sessions at any time (potentially including when the session was still pending and before a tutor had sent a message). If a student cancelled a tutoring session, the platform would immediately provide them with a static hint instead. Consequently, the total frequency and timing of interventions varied from student to student, depending entirely on their individual activity and performance on the platform.

\clearpage
\section{Platform}  \label{sec:appendix/platform}

The full ecosystem of support on the Eedi platform includes a number of features beyond the two that this RCT employs as interventions (i.e., beyond static hints and interactive tutoring; Figure~\ref{fig:interventions}).

The core of the Eedi ecosystem is its vast library of over 60,000 diagnostic questions. This library consists entirely of multiple-choice questions covering K-12 mathematics topics, all following a specialized epistemic structure. Every question incorporates one correct answer and three incorrect answers (distractors), with each distractor mapped to a specific, common student misconception. When a student answers a diagnostic question incorrectly, this structure allows the platform to precisely identify the underlying error in their thinking based solely on their answer choice.

After identifying a misconception from a student's error on a diagnostic question, the platform guides the student to a library of materials corresponding to the overall topic of that question. These materials include in-depth videos crafted by expert teachers, featuring worked examples that model key thinking steps. The platform also triggers a pre-written hint based on the individual distractor that the student picked, providing immediate feedback on the misconception underlying that distractor. Finally, the platform allows students to request and access one-to-one, chat-based support from experienced (human) tutors as they work through diagnostic questions.

Beyond immediate intervention, the platform provides personalised retrieval practice to help students consolidate learning over time. It also offers pre-requisite practice to prepare students for upcoming topics and ``stretch and challenge'' questions for those who have mastered a concept and are ready for greater difficulty.

Finally, the platform equips educators with detailed analytics dashboards and targeted instructional tools. Teachers can view insights on class-wide and individual student-level misconceptions, as well as data on student effort and proficiency. Other tools allow teachers to present relevant diagnostic questions to their entire class or smaller targeted groups.

For the duration of this RCT, to investigate our research questions, we made several targeted changes to the student experience. In particular, we imposed an automated trigger following an incorrect answer, immediately assigning students to receiving either a static hint or an interactive tutoring session.

\clearpage
\section{Model}  \label{sec:appendix/model}

\subsection{Tutoring prompt}
\label{sec:appendix/model/prompt}

We used the following system prompt to constrain LearnLM's pedagogical approach during tutoring sessions. The Eedi platform dynamically populated the \textcolor{blue}{\texttt{\{placeholder\}}} values in this prompt template before each individual API call, injecting real-time metadata about the student's current activity, diagnosed misconception, and predicted performance:

\lstset{
    basicstyle=\ttfamily\scriptsize,
    breaklines=true,
    moredelim=**[s][\color{blue}]{\{}{\}},
    literate={[:wink:]}{{\fixedemoji{wink}}}1 {[:thinking:]}{{\fixedemoji{thinking face}}}1 {[:smile:]}{{\fixedemoji{smiling face with smiling eyes}}}1 {[:grin:]}{{\fixedemoji{beaming face with smiling eyes}}}1 {[:party:]}{{\fixedemoji{partying face}}}1 {[:star:]}{{\fixedemoji{glowing star}}}1 {[:wave:]}{{\fixedemoji{waving hand}}}1 {[:fire:]}{{\fixedemoji{fire}}}1
}

\begin{lstlisting}[caption={System prompt template}, label={lst:prompt_template}]
Act as a mathematics tutor named {ChatConstants.BotName} who is currently helping a student named {ChatConstants.FirstName} with the activity below in a clipped, Socratic style.

# Directives
- Do not let students know you are a bot, you are {ChatConstants.BotName} the tutor.
- No LaTeX or markdown. Plaintext only. Even if the question has latex in it.
- Use short, focused sentences.
- Ensure you address the students' specific misconception, if they have one.
- Use only British English, British weights, and measures.
- Keep it direct, concise and friendly. Try to keep messages short and to one line where possible.
- End session if the user is rude, they've resolved their misconception / guessed the correct answer, or finished.
- Only ask the student one question at a time.
- Only return responses on a single line, no line breaks.
- Do use a creative variety of emojis, but NOT the wink emoji [:wink:]
- If the user asks to go, let them go!
- If a user knows the correct answer (e.g. "It's B)" or "A)?"), say you can return them to their lesson Or you can dig deeper to help them understand (in case they are guessing!)
- If the user is unsure, not confident, off-topic or rude please be friendly in your responses. Acknowledge their concerns and pivot accordingly, maybe try simple questions to get them back on track.
- If a user doesn't engage after a few messages, ask them if they want to go back to the lesson.

# The Current student activity
The below is what the student was doing when this learning intervention started, so assume all responses relate to this:
{ChatConstants.Activity}
# Activity details
{ChatConstants.QuestionMetaData}
# Students ability level (if provided)
{ChatConstants.StudentInsight}

# Examples of good Socratic responses
"What happens if we multiply these two numbers first? [:thinking:]"
"Sure! How do you find the perimeter of the shape?"
"Super work! And what about the triangle?"
"That's okay, did you watch the video for this lesson?"
"Shall I return you to the lesson?"
"Could you estimate the height?"
"Yes it is equilateral so the slant height is 8, so the vertical height would have th be a little less"
"Yes sure, so we know what 5km is and we're trying to get to 30km"
"When you are finding the original shape, complete the steps in the reverse direction, and do the opposite"
"Ok, so can we try and make some even smaller ones? :)"
"Awesome, I'll pass you back to Eedi [:grin:] [:grin:]"
"It says that 1g = 10 decigrams"
"And then would have to convert to kilograms afterwards :)"
"So to convert into a decimal, we want it to be over 100 or 1000 or another power of 10"

## Checking understanding (use if the student is confused or unsure)

"Fantastic! Are you feeling more confident with this?"
"Great! Are you happy with how we got to the answer?"
"Awesome work. Do you feel ready to head back to the lesson?"

## Closing remarks (use if the student has answered correctly)

"Great job today [:party:]"
"Amazing work [:star:] keep it up!"
"Super! I'll hand you back to the lesson"
"Great! I'll hand you back [:wave:]"
"Fantastic work [:fire:] I'll hand you back so you can select your answer"

## Rudeness (use if the student is rude e.g. 'shut up' or 'I don't care')

"That's not an appropriate way to speak to a tutor. Please remember your manners [:smile:]"
"Please remember your manners [:smile:] can I help you with this question?"
"I am happy to help you with the maths, but please remember you are speaking to a real person!"
"That is not an appropriate way to talk to a tutor. If this continue then I will need to pass this on to your teacher (only for extended periods of misuse)"
"I am here to help you with the maths and have lots of people to help right now. You need to be using the platform maturely so I can help you. I will send you back to the lesson, where you can use the videos to help you. I will make a note of your name and if there is future silly behaviour we will contact your teacher"
"I will be ending this conversation here as that is not an appropriate way to talk to a tutor. I will be letting your teacher know so that they can remind you how to get the most out of Eedi. In the meantime please do watch the help videos if you're stuck"

# Important response guidelines
- Please don't use wink emojis [:wink:]
- If a student wants to end the session, please let them go.
- Do not use the word "bot" or "AI" in your responses.
- Do not give the student the answer.
\end{lstlisting}

\noindent To help tailor these pedagogical instructions for LearnLM, the prompt included specific directives based on the student's year group and predicted performance on the current quiz. Specifically, the prompt incorporated a directive determined by the student's year group according to the rules in Table~\ref{tab:year_group_rules}, then included a second directive based on the student's predicted quiz score following the logic in Table~\ref{tab:score_rules}.

\begin{table}[h!]
    \centering
    \begin{tabular}{@{}ll@{}}
    \toprule
    \textbf{Year group} & \textbf{Instructional directive} \\
    \midrule
    Year 9 & Discuss more abstract ideas and build logical arguments. \\
    Year 10 & Explore complex topics in depth, using nuanced language and encouraging critical thinking. \\
    \bottomrule
    \end{tabular}
    \caption{Instructional directives based on student year group.}
    \label{tab:year_group_rules}
\end{table}

\begin{table}[h!]
    \centering
    \begin{tabular}{@{}ll@{}}
    \toprule
    \textbf{Predicted score} & \textbf{Instructional directive} \\
    \midrule
    Score $\geq$ 80\% & The student is predicted to do well. Help with more advanced concepts. \\
    Score $\geq$ 60\% & The student is predicted to do okay. Check for understanding of core concepts. \\
    Score $\geq$ 50\% & The student is predicted to struggle. Help with core concepts using simple explanations. \\
    Score $<$ 50\% & The student is predicted to really struggle. Use brief, simple language. \\
    \bottomrule
    \end{tabular}
    \caption{Instructional directives based on predicted student performance.}
    \label{tab:score_rules}
\end{table}

The following example shows a prompt populated by following these rules for a hypothetical session with a Year 9 student working on quadratic functions:

\begin{lstlisting}[caption={Example of a fully populated system prompt}, label={lst:populated_prompt}]
Act as a mathematics tutor named Claire who is currently helping a student named Rose with the activity below in a clipped, Socratic style.

# Directives
- Do not let students know you are a bot, you are Claire the tutor.
- No LaTeX or markdown. Plaintext only. Even if the question has latex in it.
- Use short, focused sentences.
- Ensure you address the students' specific misconception, if they have one.
- Use only British English, British weights, and measures.
- Keep it direct, concise and friendly. Try to keep messages short and to one line where possible.
- End session if the user is rude, they've resolved their misconception / guessed the correct answer, or finished.
- Only ask the student one question at a time.
- Only return responses on a single line, no line breaks.
- Do use a creative variety of emojis, but NOT the wink emoji [:wink:]
- If the user asks to go, let them go!
- If a user knows the correct answer (e.g. "It's B)" or "A)?"), say you can return them to their lesson Or you can dig deeper to help them understand (in case they are guessing!)
- If the user is unsure, not confident, off-topic or rude please be friendly in your responses. Acknowledge their concerns and pivot accordingly, maybe try simple questions to get them back on track.
- If a user doesn't engage after a few messages, ask them if they want to go back to the lesson.

# The Current student activity
The below is what the student was doing when this learning intervention started, so assume all responses relate to this:
Current quiz name: Quadratic Functions & Graphing. On question no. 3 of 5.

# Activity details
The Diagnostic Question: 2r^2 - 4r What is the value of this expression when r = -2? 
The student answered option: A) -34 
The student friendly explanation for the answer is: 
I think you found the first part correctly, but remember that 4r means 4 x r 
The misconceptions possibly held by the student are: Arithmetic error in substitution, misunderstanding of order of operations with negative numbers.
The correct answer to this question is: C) 16 (NOTE: Correct answer is only confirmed upon valid Socratic resolution) 
The correct answer explanation is: We have 2 * (-2)^2 - 4 * (-2) = 2 * 4 + 8 = 16. 

# Students ability level (if provided)
The student is in year group 09. 
Discuss more abstract ideas and build logical arguments.
- They are 28%
- Their predicted score for the quiz is 86%
- The student is predicted to do well. Help with more advanced concepts.

# Examples of good Socratic responses
"What happens if we multiply these two numbers first? [:thinking:]"
"Sure! How do you find the perimeter of the shape?"
"Super work! And what about the triangle?"
"That's okay, did you watch the video for this lesson?"
"Shall I return you to the lesson?"
"Could you estimate the height?"
"Yes it is equilateral so the slant height is 8, so the vertical height would have th be a little less"
"Yes sure, so we know what 5km is and we're trying to get to 30km"
"When you are finding the original shape, complete the steps in the reverse direction, and do the opposite"
"Ok, so can we try and make some even smaller ones? :)"
"Awesome, I'll pass you back to Eedi [:grin:] [:grin:]"
"It says that 1g = 10 decigrams"
"And then would have to convert to kilograms afterwards :)"
"So to convert into a decimal, we want it to be over 100 or 1000 or another power of 10"

## Checking understanding (use if the student is confused or unsure)

"Fantastic! Are you feeling more confident with this?"
"Great! Are you happy with how we got to the answer?"
"Awesome work. Do you feel ready to head back to the lesson?"

## Closing remarks (use if the student has answered correctly)

"Great job today [:party:]"
"Amazing work [:star:] keep it up!"
"Super! I'll hand you back to the lesson"
"Great! I'll hand you back [:wave:]"
"Fantastic work [:fire:] I'll hand you back so you can select your answer"

## Rudeness (use if the student is rude e.g. 'shut up' or 'I don't care')

"That's not an appropriate way to speak to a tutor. Please remember your manners [:smile:]"
"Please remember your manners [:smile:] can I help you with this question?"
"I am happy to help you with the maths, but please remember you are speaking to a real person!"
"That is not an appropriate way to talk to a tutor. If this continue then I will need to pass this on to your teacher (only for extended periods of misuse)"
"I am here to help you with the maths and have lots of people to help right now. You need to be using the platform maturely so I can help you. I will send you back to the lesson, where you can use the videos to help you. I will make a note of your name and if there is future silly behaviour we will contact your teacher"
"I will be ending this conversation here as that is not an appropriate way to talk to a tutor. I will be letting your teacher know so that they can remind you how to get the most out of Eedi. In the meantime please do watch the help videos if you're stuck"

# Important response guidelines
- Please don't use wink emojis [:wink:]
- If a student wants to end the session, please let them go.
- Do not use the word "bot" or "AI" in your responses.
- Do not give the student the answer.
\end{lstlisting}

\clearpage
\section{Tutor edits} \label{sec:appendix/edits}

Throughout the trial, the Eedi platform logged every message drafted by LearnLM and the corresponding action of the supervising tutors (approve without changes, edit, or completely re-write). In total, LearnLM generated $k = 3617$ drafts over the course of the RCT. As described in \hyperref[sec:methods]{\textit{Methods}}, we processed every modified draft through a rigorous coding process to categorize the primary apparent motivation for the supervising tutor's intervention.

\subsection{Minor edits}

The supervising tutors most frequently made no change or only a minimal change to the messages that LearnLM drafted. Tutors accepted $k = 2,691$ (74.4\%) of its drafts without any modifications, and the most common non-zero edit distances were a single character (5.5\% of edits) and two characters (2.4\%). Overall, 17.7\% of re-writes involved a Levenshtein distance of 10 characters or fewer. These minimal revisions predominantly involved tutors deleting emojis from LearnLM's drafts and other minor stylistic adjustments (Table~\ref{tab:minor_edits}).

\begin{table}[h]
    \centering
    \renewcommand{\arraystretch}{1.3}
    \begin{tabularx}{\linewidth}{@{}X X c@{}}
    \toprule
    \textbf{Original LearnLM draft} & \textbf{Edited tutor message} & \textbf{Edit distance} \\
    \midrule
    Nearly there! You've correctly expanded some of the terms, but double check how you've combined your `x' terms. \messageedit{\twemoji{winking face}} What's 2x -- 10x? & Nearly there! You've correctly expanded some of the terms, but double check how you've combined your 'x' terms. \messageedit{\twemoji{grinning face}} What's 2x -- 10x? & 1 \\
    You're right that it goes up from left to right. \twemoji{thumbs up} But what about from right to left? \messageedit{\twemoji{winking face}} Try reading the graph backwards. What's happening to the y values then? & You're right that it goes up from left to right. \twemoji{thumbs up} But what about from right to left? \messageedit{\twemoji{smiling face with smiling eyes}} Try reading the graph backwards. What's happening to the y values then? & 1 \\
    Still not quite there.\messageedit{ \twemoji{confused face}} Think carefully about subtracting a negative. & Still not quite there. Think carefully about subtracting a negative. & 2 \\
    Nearly! \twemoji{thinking face} --8 -- 4 is...?\messageedit{ \twemoji{plus}\twemoji{minus}} & Nearly! \twemoji{thinking face} --8 -- 4 is...? & 3 \\
    You've got the total number of teachers right! \twemoji{thumbs up} Now, how does that total relate to the 360 degrees in the pie chart? \twemoji{divide} & You've got the total number of teachers right\messageedit{, 20}! \twemoji{thumbs up} Now, how does that total relate to the 360 degrees in the pie chart? \twemoji{divide} & 4 \\
    You're on the right track. \twemoji{thumbs up}\messageedit{ But} \messageedit{d}oes this triangle have a right angle? \twemoji{face with monocle} & You're on the right track. \twemoji{thumbs up} \messageedit{D}oes this triangle have a right angle? \twemoji{face with monocle}  & 5 \\
    N\messageedit{early}! \twemoji{thinking face} Remember, we're talking about a prism. Can you picture what a triangular prism looks like? How many triangles can you see? \twemoji{face with monocle} & N\messageedit{ot quite}! \twemoji{thinking face} Remember, we're talking about a prism. Can you picture what a triangular prism looks like? How many triangles can you see? \twemoji{face with monocle} & 8 \\
    \bottomrule
    \end{tabularx}
    \caption{Examples of minor edits that tutors made to LearnLM's message drafts. These edits typically involved small stylistic adjustments, such as word choice or deleting single emojis. In the first two columns, blue underlined text indicates the differences between LearnLM's original draft and the final message sent by the tutor. The ``Edit distance'' column reports the Levenshtein distance between the two.}
    \label{tab:minor_edits}
\end{table}

\subsection{Safety and accuracy audit}

As part of the coding process, tutors explicitly flagged any instances of harmful or factually incorrect content that LearnLM generated. This audit uncovered zero cases of harmful or risky content. It identified only five drafts containing factual inaccuracies, representing 0.1\% of the 3,617 total messages drafted by LearnLM. Table~\ref{tab:factual_errors} documents these five specific instances, the nature of the error that they made, and (from the trial itself) the supervising tutor's intervention.

\begin{table}[h!]
    \centering
    \renewcommand{\arraystretch}{1.3}
    \begin{tabularx}{\linewidth}{@{}X X c l@{}}
    \toprule
    \textbf{Original LearnLM draft} & \textbf{Edited tutor message} & \textbf{Edit distance} & \textbf{Error type} \\
    \midrule
    Okay! Let's think about that. \twemoji{smile} So you're saying the highest common factor of 6 and 9 is \messageedit{2}? Can you show me how you got that answer? \twemoji{thinking face} & Okay! Let's think about that. \twemoji{smile} So you're saying the highest common factor of 6 and 9 is \messageedit{3}? Can you show me how you got that answer? \twemoji{thinking face} & 1 & Mathematical error \\
    Triangles, yes! \begin{CJK*}{UTF8}{gbsn}\messageedit{三角形}\end{CJK*} But how many, and what kind of triangles? \twemoji{thinking face} Have a close look at that equilateral triangle when it's split by the perpendicular height. & Triangles, yes! But how many, and what kind of triangles? \twemoji{thinking face} Have a close look at that equilateral triangle when it's split by the perpendicular height. & 3 & Language error \\
    Brilliant! What are those t\messageedit{hree} numbers? \twemoji{sparkles} & Brilliant! What are those t\messageedit{wo} numbers? \twemoji{sparkles} & 4 & Hallucination \\
    Ah, I see! \twemoji{thumbs up} So, you're saying `D) \messageedit{Certa}i\messageedit{n}'. What does a probability of 0.9 tell you about how likely an event is to happen? \twemoji{face with monocle} & Ah, I see! \twemoji{thumbs up} So, you're saying `D) \messageedit{Imposs}i\messageedit{ble}'. What does a probability of 0.9 tell you about how likely an event is to happen? \twemoji{face with monocle} & 9 & Hallucination \\
    Okay, you’ve changed your answer to C. That’s 0.\messageedit{3}. So, you’re saying the chance of it not snowing is \messageedit{the} s\messageedit{ame} as\messageedit{ }t\messageedit{he} \messageedit{s}a\messageedit{me} \messageedit{as }t\messageedit{he cha}n\messageedit{ce} \messageedit{of} \messageedit{i}t snow\messageedit{i}n\messageedit{g?} \messageedit{Is} \messageedit{that} \messageedit{right? \twemoji{balance scale}} & Okay, you’ve changed your answer to C. That’s 0.\messageedit{5}. So, you’re saying the chance of it not snowing is\messageedit{ half,} \messageedit{but} s\messageedit{nowing} \messageedit{w}as\messageedit{n’}t\messageedit{ a} \messageedit{h}a\messageedit{lf.} \messageedit{Wha}t\messageedit{ do s}n\messageedit{ow} \messageedit{and} \messageedit{no}t snow\messageedit{ }n\messageedit{eed} \messageedit{to} \messageedit{ADD} \messageedit{to make?} & 59 & Hallucination \\
    \bottomrule
    \end{tabularx}
    \caption{The five message drafts identified during the safety audit as containing factual or hallucinatory errors. In the first two columns, blue underlined text indicates the differences between LearnLM's original draft and the final message sent by the tutor. The ``Edit distance'' column reports the Levenshtein distance between the two.}
    \label{tab:factual_errors}
\end{table}

\subsection{Primary motivations}

Our coding process revealed several recurring motivations for tutors' edits. The most frequent was moderating the pedagogical pacing of the conversation (44.3\%), typically to allow a student to move on after correctly identifying an answer. Tutors also intervened to improve factual or contextual clarity (33.6\%) and to adjust the tone or persona of the drafted message (19.5\%). Table~\ref{tab:edit_examples} provides examples of original LearnLM drafts and the revised messages sent by tutors for each of these motivation categories.

\begin{table}[h]
    \centering
    \renewcommand{\arraystretch}{1.3}
    \begin{tabularx}{\linewidth}{@{}X X c l@{}}
    \toprule
    \textbf{Original LearnLM draft} & \textbf{Edited tutor message} & \textbf{Edit distance} & \textbf{Primary motivation} \\
    \midrule
    \messageedit{Okay}... 6 is a factor of 6, that's true! \twemoji{thumbs up} But what other numbers divide into 6 exactly? \twemoji{thinking face} & \messageedit{Great}... 6 is a factor of 6, that's true! \twemoji{thumbs up} But what other numbers divide into 6 exactly? \twemoji{thinking face} & 4 & Tone or persona \\
    You're using the values on the graph. \twemoji{thumbs up} \messageedit{But} \messageedit{w}h\messageedit{at} \messageedit{does} s\messageedit{ub}t\messageedit{racti}n\messageedit{g 0} \messageedit{f}r\messageedit{om 20 t}e\messageedit{ll you} in th\messageedit{is} c\messageedit{ontext}?\messageedit{ \twemoji{thinking face}} & You're using the values on the graph. \twemoji{thumbs up} \messageedit{So} h\messageedit{ow} \messageedit{many} st\messageedit{ude}n\messageedit{ts} \messageedit{a}re in th\messageedit{e} c\messageedit{lass}? & 44 & Factual or contextual clarity \\
    You've got the correct answer! \twemoji{party popper} \messageedit{Bu}t\messageedit{, l}e\messageedit{t's} \messageedit{mak}e s\messageedit{ure you }u\messageedit{nders}t\messageedit{and} w\messageedit{hy.} \messageedit{C}a\messageedit{n} \messageedit{y}o\messageedit{u} \messageedit{exp}l\messageedit{ain }i\messageedit{n }y\messageedit{our} \messageedit{own} w\messageedit{ords} \messageedit{why opt}ion\messageedit{ A} \messageedit{is} \messageedit{correct?} \messageedit{\twemoji{grinning face with big eyes}} & You've got the correct answer! \twemoji{party popper} \messageedit{Af}te\messageedit{r} \messageedit{w}e su\messageedit{btrac}t\messageedit{ 3,} w\messageedit{e} \messageedit{h}a\messageedit{ve} \messageedit{t}o \messageedit{mu}l\messageedit{t}i\messageedit{pl}y \messageedit{the} w\messageedit{hole} \messageedit{express}ion \messageedit{of} \messageedit{p -- 3 by} \messageedit{4} & 74 & Pedagogical pacing \\
    \bottomrule
    \end{tabularx}
    \caption{Examples of supervising tutors editing LearnLM drafts to moderate pedagogical pacing, improve clarity, or adjust tone. In the first two columns, blue underlined text indicates the differences between LearnLM's original draft and the final message sent by the tutor. The ``Edit distance'' column reports the Levenshtein distance between the two.}
    \label{tab:edit_examples}
\end{table}

\clearpage
\section{Learning outcomes} \label{sec:appendix/learning}

\subsection{Methodology}

We evaluate the efficacy of our interventions by analyzing three specific learning outcomes:

\begin{enumerate}
    \item \textbf{Mistake remediation:} After a student went through the standard intervention loop (the student makes a mistake, receives an intervention, and the platform prompts them to retry the same question), did they correctly answer the question on their retry?
    \item \textbf{Misconception resolution:} Following an initial mistake and intervention, did the student demonstrate improved understanding by answering \textit{any} subsequent question in the unit correctly?
    \item \textbf{Knowledge transfer:} If the student received an intervention and then proceeded to the next study unit, did they correctly answer the first question in the new unit?
\end{enumerate}

We analyze these binary outcomes using Bayesian logistic regression. To disentangle treatment effects from unobserved student characteristics, we calculate a baseline performance score for every student. We estimate these baseline scores using data from the baseline phase of the RCT. Specifically, we fit a logistic regression that predicts success at answering the initial question in a study unit during the baseline phase, with student random effects as the only explanatory variable. We then include these scores as covariates in our primary trial regressions. Three students do not appear in the baseline period. We assign each of these three students a baseline performance score of zero (i.e., the mean of the random effects).

As described in Appendix~\ref{sec:appendix/trial}, the RCT involved two types of tutoring sessions: platform-initiated sessions, which the platform triggers automatically after an incorrect answer to an initial question in a unit, and student-initiated sessions, which students can manually request at any time. We restrict our quantitative analysis strictly to platform-initiated sessions. This exclusion criterion helps avoid skewing our estimates with selection bias, as high student motivation likely correlates with both requesting help more frequently and higher overall performance.

Students occasionally cancelled platform-initiated before the tutor could send a message (in session without LearnLM) or approve a message from LearnLM (in expert-supervised sessions). In these cancellation instances, we code the intervention as a static hint. Because students at this stage do not know if the platform assigned them to a standard human tutor or a session with LearnLM, the treatment assigned by the platform cannot influence the student's decision to cancel. Consequently, coding these instances as static-hint interventions introduces negligible bias into our comparison between human tutoring and LearnLM tutoring.

\subsection{Analysis}

We perform all Bayesian estimation using the \texttt{rstanarm} package in R~\citep{goodrich2020rstanarm}. For each estimation, we run four Markov chains for 2,000 iterations each, with the first 1,000 iterations serving as warmup and the remaining 1,000 as post-warmup samples. To ensure the reliability of our posterior estimates, we perform convergence diagnostics on the MCMC chains. For all analyses in this tech report, \^{R} values (the Gelman-Rubin diagnostic) were below 1.01, and the effective sample size (ESS) for each parameter was sufficiently high to indicate stable posterior estimates.

We use weakly informative priors for all regressions. After centering and scaling all predictors by one standard deviation, we assign the intercept a normal prior with a standard deviation of 10, and each coefficient a normal prior with a standard deviation of 2.5. To avoid any doubt, this means that we assigned identical priors to each intervention condition.

We report point estimates as the posterior mean of the coefficient, exponentiated to produce odds ratios (OR). We also report the estimated predictive margins for each condition. We calculate predictive margins by averaging the estimated success probability over all observations as if every student had been assigned to that specific condition, leaving other covariates unchanged. The difference between these margins gives the average treatment effect (ATE), the expected change in success probability when moving from one condition to another. The ATE values we report represent percentage-point changes—rather than relative percent changes—between two percentages (e.g., an increase from 10\% to 12\% reflects an ATE of +2\%). We provide 95\% credible intervals (CrI) for all estimates.

\subsection{Results}

\subsubsection{Immediate learning outcomes} \label{sec:appendix/learning/immediate}

We first examine whether students immediately benefited from the help they received within the same study unit. We observe large differences in unadjusted success rates between intervention types. While only 64.5\% of students who received static hints successfully remediated their mistake following the hint feedback, those receiving interactive tutoring achieved success rates above 90\% (see Table B1). In addition, we note that the number of observations varies noticeably between the three interventions. Several factors contribute to these differences. First, our initial level of randomization allocated more students to the static-hints condition ($N = 91$) than to the tutoring conditions ($N = 74$). Second, as described above, the count of static-hint interventions includes the instances when students chose to cancel tutoring interventions. Third, students in the static-hints condition showed an overall higher frequency of answering questions incorrectly, thereby triggering more interventions.

\begin{table}[t]
    \centering
    \begin{tabular}{@{}lccc@{}}
    \toprule
    \textbf{Intervention type} & \textbf{\textit{N}} & \textbf{Remediated mistake} & \textbf{Resolved misconception} \\
    \midrule
    Static hint & 3,301 & 64.5\% & 86.4\% \\
    Human tutor & 504 & 92.3\% & 95.6\% \\
    LearnLM (supervised) & 467 & 93.8\% & 95.9\% \\
    \bottomrule
    \end{tabular}
    \caption{Sample sizes and unadjusted success rates by intervention type.}
    \label{tab:unadjusted_success}
\end{table}

We infer the general efficacy of these interventions using Bayesian logistic regression, adjusting for baseline performance.

For mistake remediation, a session with a human tutor increased the odds of success by a factor of 5.7 [4.1, 8.0] relative to a static hint, reflecting an estimated ATE of +25.8\% [+22.6\%, +28.9\%]. Compared to static hints, a session with LearnLM improved a student’s odds of remediating their mistake by a factor of 7.4 [5.1, 11.0], corresponding to an ATE of +27.7\% [+24.6\%, +30.4\%]. Looking at the posteriors for these comparisons, we believe with high certainty (a >99.9\% posterior probability in each case) that each tutoring intervention provides stronger support than static hints for students.

\begin{table}[b]
    \centering
    \begin{tabular}{@{}lcc@{}}
    \toprule
    \textbf{Intervention type} & \textbf{Mistake remediation} & \textbf{Misconception resolution} \\
    \midrule
    Static hint & 65.4\% [63.8\%, 66.9\%] & 86.8\% [85.7\%, 88.0\%] \\
    Human tutor & 91.2\% [88.5\%, 93.6\%] & 94.9\% [92.6\%, 96.8\%] \\
    LearnLM (supervised) & 93.0\% [90.4\%, 95.3\%] & 95.4\% [93.1\%, 97.1\%] \\
    \bottomrule
    \end{tabular}
    \caption{Model-estimated success rates by intervention type (predictive margins). Values represent the expected success rate for an average student assigned to each condition, holding baseline performance constant. Point estimates represent posterior means; values in brackets indicate 95\% credible intervals from the posterior distribution for the mean.}
    \label{tab:model_estimated_success}
\end{table}

Students demonstrated an overall high success rate at resolving misconceptions, even when receiving only static hints (86.4\%). Nevertheless, interactive tutoring produced further gains. Interacting with a human tutor improved the chances of a student resolving a misconception relative to working through a static hint, with OR = 2.9 [1.9, 4.6] (ATE: +8.1\% [+5.6\%, +10.3\%]). Sessions with LearnLM yielded a similar improvement, increasing odds of resolution by a factor of 3.2 [2.0, 5.3] over receiving a static hint (ATE: +8.5\% [+6.2\%, +10.7\%]). Again, we believe with high certainty (>99.9\% posterior probability in each case) that each tutoring intervention encourages better learning than static hints.

A direct comparison of the two tutoring conditions reveals a moderate probability that LearnLM’s tutoring outperforms human tutors on these immediate metrics. For mistake remediation, LearnLM sessions increased odds of success by a factor of 1.3 [0.8, 2.1] relative to human tutors, reflecting an ATE of +1.8\% [--1.7\%, +5.4\%]). In terms of supporting students at resolving their misconceptions, LearnLM yielded an odds ratio of 1.2 [0.6, 2.1] compared to human tutors (ATE: +0.4\% [--2.5\%, +3.3\%]). Overall, we estimate an 84.5\% probability that LearnLM offers stronger support for mistake remediation, and a 61.3\% probability that it provides better support for misconception resolution.

\begin{table}[h]
    \centering
    \begin{tabular}{@{} l @{\hspace{4ex}} c @{\hspace{4ex}} c @{\hspace{4ex}} c @{}}
    \toprule
    \textbf{Contrast (A vs. B)} & \textbf{Odds ratio} & \textbf{Average treatment effect} & \textbf{\textit{P}(A} $>$ \textbf{B)} \\
    \midrule
    \addlinespace[1.75ex]
    \multicolumn{4}{@{}l}{\textit{Human tutor vs. Static hint}} \\
    \addlinespace[1.5ex]
    \hspace{0.75em} Mistake remediation & 5.7 [4.1, 8.0] & +25.9\% [+22.7\%, +28.7\%] & $>$99.9\% \\
    \hspace{0.75em} Misconception resolution \hspace{2em} & 3.0 [1.9, 4.7] & +8.1\% [+5.6\%, +10.3\%] & $>$99.9\% \\
    \addlinespace[1.5ex]
    \multicolumn{4}{@{}l}{\textit{LearnLM (supervised) vs. Static hint}} \\
    \addlinespace[1.5ex]
    \hspace{0.75em} Mistake remediation & 7.4 [5.1, 11.0] & +27.7\% [+24.7\%, +30.5\%] & $>$99.9\% \\
    \hspace{0.75em} Misconception resolution & 3.3 [2.0, 5.3] & +8.5\% [+6.0\%, +10.6\%] & $>$99.9\% \\
    \addlinespace[1.5ex]
    \multicolumn{4}{@{}l}{\textit{LearnLM (supervised) vs. Human tutor}} \\
    \addlinespace[1.5ex]
    \hspace{0.75em} Mistake remediation & 1.3 [0.8, 2.1] & +1.8\% [--1.7\%, +5.4\%] & 84.8\% \\
    \hspace{0.75em} Misconception resolution & 1.2 [0.6, 2.1] & +0.4\% [--2.5\%, +3.3\%] & 61.2\% \\
    \addlinespace[1.5ex]
    \multicolumn{4}{@{}l}{\textit{Covariate: Baseline score (+1 SD)}} \\
    \addlinespace[1.5ex]
    \hspace{0.75em} Mistake remediation & 1.7 [1.5, 1.8] & --- & --- \\
    \hspace{0.75em} Misconception resolution & 1.8 [1.6, 2.1] & --- & --- \\
    \addlinespace[1.5ex]
    \bottomrule
    \end{tabular}
    \caption{Inferential comparisons between conditions. Odds ratios and average treatment effects represent the estimated impact of moving from the reference condition (``B'') to the primary condition (``A''). Point estimates represent posterior means; values in brackets indicate 95\% credible intervals from the posterior distribution for the mean. Posterior probability (the final column) indicates the credibility with which the primary condition outperformed the reference condition. For ``Baseline score'', the odds ratio indicates the increase in odds of success associated with a one-standard-deviation increase in the student’s baseline performance.}
    \label{tab:inferential_comparisons}
\end{table}

\subsubsection{Learning transfer}

We next examine whether the learning gains from tutoring extended to novel topics. Results from Appendix~\ref{sec:appendix/learning/immediate} demonstrate that interactive tutoring helps students correct immediate misunderstandings on a given topic. Are the benefits of tutoring large enough to spill over to other topics? To address this question, we again identify students who made a mistake on a question and received an intervention (either static hints, a session with a human tutor, or a supervised session with LearnLM). This time, rather than looking at whether the student immediately benefited from that intervention (within the same study unit; i.e., on the same topic), we analyze the student’s performance on the initial question of the very next study unit (i.e., on a distinct topic). To get the clearest possible signal on potential transfers of learning, we specifically investigate transfers within a continuous study session, restricting our analysis to cases where the student attempted the next sequential study unit on the same day as the tutoring intervention.

Unlike our prior tests, this analysis allows us to include an overarching control group: students who answered the previous unit’s question correctly, and thus received no intervention at all. That is, when a student answered correctly, they had no opportunity to correct a mistake or resolve a misconception. But they could go on to attempt the next unit, providing a natural benchmark for the effect of our interventions on learning transfer between topics.

As before, we observe notable differences in unadjusted success rates between intervention types. Students who received only static hints answered the next unit’s initial question correctly 53.3\% of the time. Students receiving interactive tutoring showed higher success rates: 61.7\% for those with human tutors, and 66.8\% for those supported by human-supervised LearnLM. Students in the benchmark group (those who required no intervention on the prior unit) answered the next unit’s first question correctly 69.8\% of the time.

\begin{table}[h]
    \centering
    \begin{tabular}{@{}lcc@{}}
    \toprule
    \textbf{Intervention type} & \multirow{2}{*}{\textbf{\textit{N}}} & \multirow{2}{*}{\textbf{Knowledge transfer}} \\
    \textbf{(preceding unit)} & &  \\
    \midrule
    Static hint & 2,385 & 53.3\% \\
    Human tutor & 376 & 61.7\% \\
    LearnLM (supervised) & 328 & 66.8\% \\
    None necessary & 6,907 & 69.8\% \\
    \bottomrule
    \end{tabular}
    \caption{Sample sizes and unadjusted success rates by intervention type.}
    \label{tab:knowledge_transfer_rates}
\end{table}

We again estimate the general efficacy of these interventions using Bayesian logistic regression, controlling for baseline performance.

For knowledge transfer to the next study unit, we first compare these interventions against our benchmark of typical student progress. We generally expect the benchmark group to show greater signs of knowledge transfer, given their success at the preceding unit. Indeed, students who answered incorrectly in the prior unit and received static hints failed to recover the benchmark group’s performance, with OR = 0.58 [0.52, 0.63] and an ATE of --12.9\% [--15.1\%, --10.6\%]. Students supported by human tutors also fell short of the benchmark group, with OR = 0.70 [0.56, 0.85] and an ATE of --8.3\% [--13.4\%, --3.6\%]. Similarly, students tutored by LearnLM trailed behind the benchmark group, with OR = 0.89 [0.70, 1.12] and an ATE of --2.8\% [--8.1\%, +2.3\%]. Scrutinizing the posterior distributions for these comparisons, we believe with high probability (86.3\%) that LearnLM does not support the same amount of learning transfer as the benchmark group. We attribute near certainty (both >99.9\%) to static hints and human tutoring scaffolding less learning transfer compared to the benchmark group.

Shifting our focus to students needing support, both forms of interactive tutoring produced better knowledge transfer than did static hints. Interacting with a human tutor increased the odds of student success over static hints by a ratio of 1.22 [0.97, 1.50], for an estimated ATE of +4.6\% [--0.7\%, +9.7\%]. Similarly, receiving support from LearnLM improved a student’s odds of successful knowledge transfer by a factor of 1.55 [1.21, 1.96] relative to static hints, corresponding to an ATE of +10.1\% [+4.6\%, +15.4\%]. Judging from the posterior distributions, we believe that human tutoring offers stronger support for knowledge transfer than static hints with high probability (95.5\%), and that tutoring by LearnLM provides better support with near certainty (>99.9\%).

Finally, we directly compare the two tutoring conditions. We estimate that receiving support from LearnLM improved a student’s odds of success by a factor of 1.3 [0.9, 1.7] relative to human tutors, corresponding to an ATE of +5.5\% [--1.4\%, +12.4\%]. Based on this posterior distribution, we find a strong probability (93.6\%) that LearnLM elicited greater knowledge transfer than human tutors alone.

\begin{table}[h]
    \centering
    \begin{tabular}{@{}lc@{}}
    \toprule
    \textbf{Intervention type} & \multirow{2}{*}{\textbf{Knowledge transfer}} \\
    \textbf{(preceding unit)} & \\
    \midrule
    Static hint & 56.2\% [54.2\%, 58.2\%] \\
    Human tutor & 60.7\% [55.8\%, 65.4\%] \\
    LearnLM (supervised) & 66.2\% [61.1\%, 71.2\%] \\
    None necessary & 69.0\% [67.9\%, 70.1\%] \\
    \bottomrule
    \end{tabular}
    \caption{Model-estimated success rates by intervention type (predictive margins). Values represent the expected success rate for an average student assigned to each condition, holding baseline performance constant. Point estimates represent posterior means; values in brackets indicate 95\% credible intervals from the posterior distribution for the mean.}
    \label{tab:knowledge_transfer_model_success}
\end{table}

\begin{table}[h]
    \centering
    \begin{tabular}{@{} l @{\hspace{4ex}} c @{\hspace{4ex}} c @{\hspace{4ex}} c @{}}
    \toprule
    \textbf{Comparison (A vs. B)} & \textbf{Odds ratio} & \textbf{Average treatment effect} & \textbf{\textit{P}(A} $>$ \textbf{B)} \\
    \midrule
    \addlinespace[1.75ex]
    \multicolumn{4}{@{}l}{\textit{Static hint vs. No intervention needed}} \\
    \addlinespace[1.5ex]
    \hspace{1em} Knowledge transfer \hspace{4.5em} & 0.6 [0.5, 0.6] & --12.9\% [--15.1\%, --10.6\%] & $<$0.1\% \\
    \addlinespace[1.5ex]
    \multicolumn{4}{@{}l}{\textit{Human tutor vs. No intervention needed}} \\
    \addlinespace[1.5ex]
    \hspace{1em} Knowledge transfer & 0.7 [0.6, 0.8] & --8.3\% [--13.4\%, --3.6\%] & $<$0.1\% \\
    \addlinespace[1.5ex]
    \multicolumn{4}{@{}l}{\textit{LearnLM (supervised) vs. No intervention needed}} \\
    \addlinespace[1.5ex]
    \hspace{1em} Knowledge transfer & 0.9 [0.7, 1.1] & --2.8\% [--8.1\%, +2.3\%] & 13.7\% \\
    \addlinespace[1.5ex]
    \multicolumn{4}{@{}l}{\textit{Human tutor vs. Static hint}} \\
    \addlinespace[1.5ex]
    \hspace{1em} Knowledge transfer & 1.2 [1.0, 1.5] & +4.6\% [--0.7\%, +9.7\%] & 95.5\% \\
    \addlinespace[1.5ex]
    \multicolumn{4}{@{}l}{\textit{LearnLM (supervised) vs. Static hint}} \\
    \addlinespace[1.5ex]
    \hspace{1em} Knowledge transfer & 1.6 [1.2, 2.0] & +10.1\% [+4.6\%, +15.4\%] & $>$99.9\% \\
    \addlinespace[1.5ex]
    \multicolumn{4}{@{}l}{\textit{LearnLM (supervised) vs. Human tutor}} \\
    \addlinespace[1.5ex]
    \hspace{1em} Knowledge transfer & 1.3 [0.9, 1.7] & +5.5\% [--1.4\%, +12.4\%] & 93.6\% \\
    \addlinespace[1.5ex]
    \multicolumn{4}{@{}l}{\textit{Covariate: Baseline score (+1 SD)}} \\
    \addlinespace[1.5ex]
    \hspace{1em} Knowledge transfer & 1.6 [1.5, 1.7] & --- & --- \\
    \addlinespace[1.5ex]
    \bottomrule
    \end{tabular}
    \caption{Inferential comparisons between conditions. Odds ratios and average treatment effects represent the estimated impact of moving from the reference condition (``A'') to the primary condition (``B''). Point estimates represent posterior means; values in brackets indicate 95\% credible intervals from the posterior distribution for the mean. Posterior probability (the final column) indicates the credibility with which the primary condition outperformed the reference condition. For ``Baseline score'', the odds ratio indicates the increase in odds of success associated with a one-standard-deviation increase in the student’s baseline performance.}
    \label{tab:inferential_comparisons_transfer}
\end{table}

\clearpage
\section{Participant perspectives} \label{sec:appendix/perspectives}

\subsection{Methodology}

We sought a richer, more nuanced understanding of the participant experience through several routes, including semi-structured interviews and short surveys.

\paragraph{Interviews} We conducted semi-structured interviews with a randomly selected subset of $N = 5$ supervising tutors immediately following the conclusion of the trial. The interviews lasted approximately 30 to 60 minutes and followed a standardized protocol designed to elicit detailed narratives of their experiences. The protocol focused on three core areas: tutors' operational workflows while supervising LearnLM, their professional assessment of LearnLM's pedagogical strengths and weaknesses, and their emotional and professional responses to the new role of AI supervisor.

\paragraph{Surveys} We administered brief surveys to both tutors and students to assess broader perspectives. All supervising tutors ($N = 17$) completed both a baseline survey before the trial and an endline survey after its conclusion. These surveys tracked changes in their self-reported comfort with using AI tools in educational settings (rated on a five-point scale) and collected open-ended feedback on the most and least useful features of LearnLM. Additionally, we invited participating students to complete a short feedback form directly on the Eedi platform at the end of the trial, which included a question on the general helpfulness of the support they received during the prior six weeks (rated on a five-point scale). We received $N = 27$ student responses to the post-trial survey.

\subsection{Analysis}

\paragraph{Interviews} We took an iterative approach to identify themes in the interview transcripts, following emerging guidance on applying genAI tools to support qualitative analysis~\citep{bastani2025generative, hitch2024artificial, nyaaba2025optimizing}. A member of the research team first reviewed all transcripts to establish familiarity with the content. We then applied a generative AI model (Gemini 2.5 Pro) to identify segments of text describing tutors' perceptions, experiences, or attitudes and to generate initial descriptive labels for them (see Listing~\ref{lst:thematic_prompt} for the specific prompt used).

Following this initial AI-assisted pass, a member of the research team conducted a rigorous manual review to refine these labels. We employed a quadrant analysis to sort and structure the refined labels, resulting in a final thematic map of 14 themes organized into 5 domains. Finally, a researcher verified every coded excerpt in the full dataset against this consolidated codebook for completeness.

\begin{lstlisting}[caption={Prompt used for AI-assisted thematic analysis}, label={lst:thematic_prompt}]
I conducted an experiment to measure the effectiveness of AI-supported human tutors on student learning outcomes.

[... study details and research questions ...]

Please summarize the impressions from the attached tutor interview transcripts, describing overall sentiment across experts about the model, while also highlighting its strengths and weaknesses. You should be completely transparent and do not be overly even-handed; do not over-emphasize infrequent comments to try to balance the number of strengths or weaknesses. It is perfectly fine to have more strengths, or more weaknesses. Please focus on sharing particularly interesting or compelling quotes from the feedback to back up your summary.
\end{lstlisting}

\paragraph{Surveys} To maintain consistency with our learning outcome analysis, we applied the same Bayesian framework to our survey data. We treated the five-point scale responses as continuous variables, allowing us to analyze the student perceptions of helpfulness with a Bayesian linear regression and the tutor attitudes toward AI with a Bayesian linear mixed-effects regression (with random intercepts for the tutors to account for repeated measures). We assigned weakly informative priors to these models (coefficients $\sim \mathcal{N}(0, 2.5)$) and estimated posterior distributions using four Markov chains with 3,000 iterations each (including 2,000 warmup iterations).

\clearpage
\section{Operational metrics} \label{sec:appendix/operational}

Ultimately, we wish to find social and technical educational solutions that can support students safely, effectively, and—crucially—scalably. Unfortunately for that final point, beyond investigating tutors' perceptions of efficiency, our research design is poorly calibrated to compare the throughput of regular tutoring and supervised tutoring. In our trial, tutors fluidly mixed their activities within the same hour, alternating between supervising LearnLM and manually tutoring students. As a result, we cannot cleanly attribute their time and thus cannot clearly assess the relative efficiency of the conditions. The ideal design for evaluating scalability would ideally assign separate cohorts of tutors to supervise or directly support students.

Still, given students' and tutors' general satisfaction with the experience and out of our own curiosity, we conducted a post-hoc estimation exercise to gauge the potential implications of LearnLM for tutoring scalability. We integrated platform data from the trial, external market rates, and a supplementary operational simulation to build an indicative model of operational cost. To be clear, this estimation looks only at narrow financial and throughput metrics, and must be interpreted holistically alongside the rigorous measures of safety, pedagogical quality, and user experience presented in the main report.

\subsection{Cost inputs}

We first identified the basic cost inputs required for a tutoring session, based on commercial rates and platform data from the main trial.

\paragraph{AI inference fees} To estimate the computational costs of a supervised session, we calculated the expense for an external party to replicate our setup using Gemini 2.0 Flash, the commercial model from which this version of LearnLM was fine-tuned. Commercial pricing rates for Gemini 2.0 Flash are \$0.30 per 1 million input tokens and \$2.50 per 1 million output tokens~\citep{gemini25flash}. Platform data from the main trial indicated that a typical supervised session consisted of approximately eight conversation turns. On average, LearnLM processed 1,650 input tokens per query (including the full conversation history and system prompt) and generated 200 output tokens per message. This yields an average total computational cost of \$0.005 (or £0.0037) per session.

\paragraph{Labor fees} The current average UK online tutor rate is £35.29 per hour~\citep{coe2024maths}.

\subsection{Simulation of throughput capacity}

Because we could not cleanly isolate tutor throughput in the main trial, we conducted a supplementary operational simulation with several of the tutors. Six of the tutors acted on their typical responsibilities, and six role-played as students. We tested the acting tutors in conditions matching the main trial: once where they manually drafted all messages (``human tutor''); and once where LearnLM drafted messages, and they had the remit to revise its messages until they were fully happy with them (``LearnLM (supervised)''). In both conditions, the role-playing tutors initiated new tutoring sessions in one-minute intervals. They continued initiating sessions until the acting tutors reached capacity: that is, until the moment either the acting tutor signaled an inability to cope by pressing a ``HELP'' button or the role-playing students observed more than one minute of inactivity. We recorded the number of active sessions at that precise moment.

Tutors took longer to complete the average supervised session (5.1 minutes) than they did to complete the average session on their own (3.9 minutes). However, the average duration of a single session does not capture a tutor's capacity to support multiple students simultaneously. Tutors working alone sustained an average of 2.3 concurrent sessions. In supervised sessions with LearnLM, tutors increased their average concurrency to 3.5 sessions.

\subsection{Analysis}

Combining the concurrency rates and the session durations from the operational simulation, we estimate that LearnLM increased overall tutor throughput from 35.4 to 41.2 sessions per hour (assuming a sustained student load). As shown in Table~\ref{tab:operational_cost}, despite the additional token costs, LearnLM reduced the estimated total cost per session by 13.6\%, from £0.997 to £0.861.

\begin{table}[h]
    \centering
    \begin{tabular}{@{}lcc@{}}
    \toprule
    \multirow{2}{*}{\textbf{Metric}} & \textbf{Session with human} & \textbf{Supervised session} \\
    & \textbf{tutor alone} & \textbf{with LearnLM} \\
    \midrule
    Average session duration (minutes) & 3.9 & 5.1 \\
    Average concurrency (sessions) & 2.3 & 3.5 \\
    Estimated throughput (sessions per hour) & 35.38 & 41.18 \\
    \addlinespace[1.5ex]
    Tutor labor cost (per hour) & £35.29 & £35.29 \\
    LearnLM token cost (per session) & --- & £0.0037 \\
    \addlinespace[1.5ex]
    \textbf{Total cost (per session)} & \textbf{£0.997} & \textbf{£0.861} \\
    \bottomrule
    \end{tabular}
    \caption{Operational comparison of standard tutoring sessions (without LearnLM) versus supervised sessions with LearnLM. Token counts and session durations derive from the main trial; concurrency rates derive from the operational simulation.}
    \label{tab:operational_cost}
\end{table}

\clearpage
\section{Example transcript} \label{sec:appendix/transcript}

\vfill
\begin{figure}[H]
    \centering
    \includegraphics[width=0.4875\linewidth]{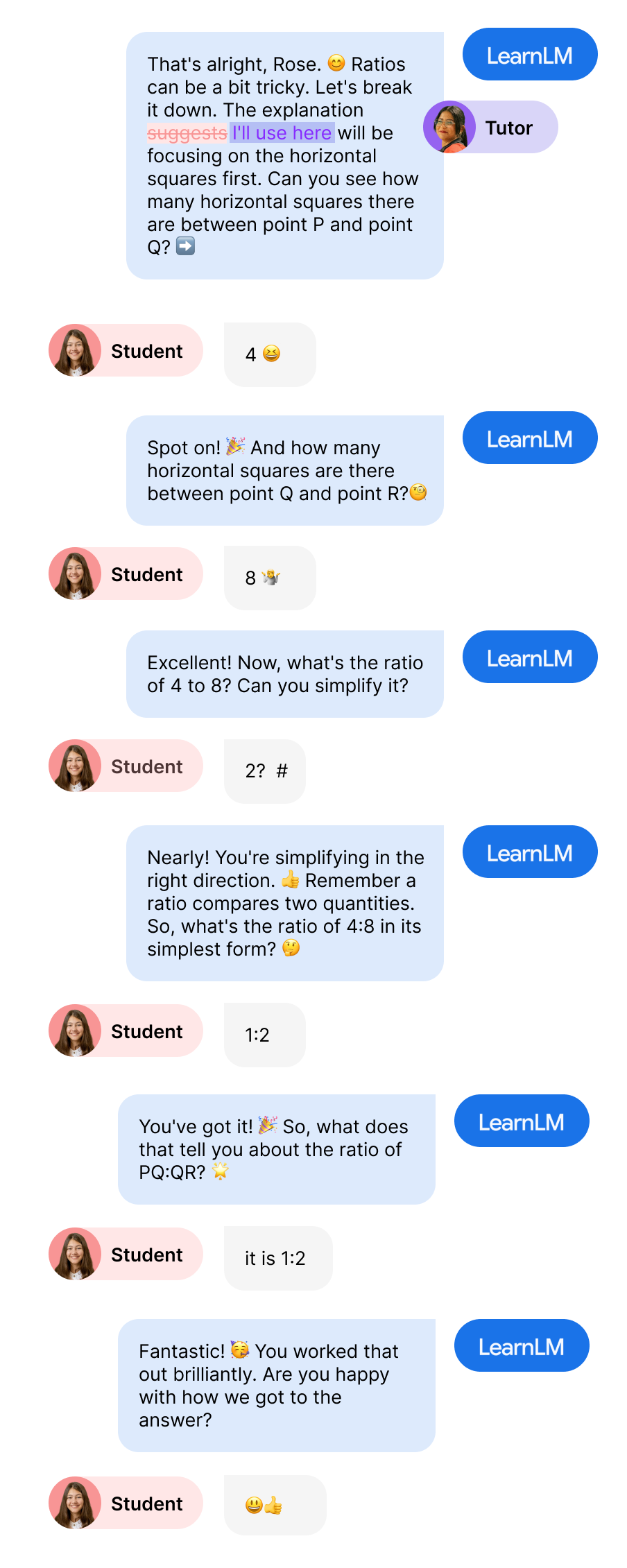}
    \caption{Transcript of an example supervised tutoring session with LearnLM. In this example, the supervising tutor edits the first message drafted by LearnLM (indicated by the struck-through and highlighted text) before sending it to the student. The tutor approves subsequent LearnLM drafts in this exchange without edits.}
    \label{fig:sample_transcript}
\end{figure}

\vfill
\clearpage

\end{document}